\documentclass[aps,prl,preprintnumbers,amsmath,amssymb,latexsym,nofootinbib,array,enumerate,letter,twocolumn,superscriptaddress]{revtex4-1}
\usepackage{here}
\usepackage{comment,braket}
\usepackage{epsf}
\usepackage{amsmath}
\usepackage{graphics}
\usepackage{amsfonts}
\usepackage{amssymb}
\usepackage{latexsym}
\usepackage{color}
\usepackage{ulem}
\input{colordvi.tex}
\usepackage{feynmp}

\usepackage[pdftex]{graphicx}
\usepackage{bm}
\usepackage[pdftex]{hyperref}
\usepackage[svgnames]{xcolor}
\definecolor{phthaloblue}{rgb}{0.0, 0.06, 0.54}
\hypersetup{
    colorlinks=true,
    linkcolor=phthaloblue,
    citecolor=blue,
    filecolor=blue,
    urlcolor=phthaloblue,
    }

\newcommand{\nc}{\newcommand}
\nc{\cH}{\mathcal{H}}
\nc{\beq}{\begin{equation}}
\nc{\eeq}{\end{equation}}

\begin{document}

\title{
Axion baryogenesis puts a new spin on the Hubble tension\\
}

\author{Raymond T.~Co}
\affiliation{William I. Fine Theoretical Physics Institute, School of Physics and Astronomy, University of Minnesota, Minneapolis, MN 55455, USA}
\affiliation{Physics Department, Indiana University, Bloomington, IN, 47405, USA}
\author{Nicolas Fernandez}
\affiliation{NHETC, Department of Physics and Astronomy, Rutgers University, Piscataway, NJ 08854, USA}
\author{Akshay~Ghalsasi}
\affiliation{
Pittsburgh Particle Physics, Astrophysics, and Cosmology Center,  Department of Physics and Astronomy, University of Pittsburgh, Pittsburgh, PA 15260, USA}
\author{Keisuke Harigaya}
\affiliation{Department of Physics, University of Chicago, Chicago, IL 60637, USA}
\affiliation{Enrico Fermi Institute and Kavli Institute for Cosmological Physics, University of Chicago, Chicago, IL 60637, USA}
\affiliation{Kavli Institute for the Physics and Mathematics of the Universe (WPI),
The University of Tokyo Institutes for Advanced Study,
The University of Tokyo, Kashiwa, Chiba 277-8583, Japan}
\author{Jessie Shelton}
\affiliation{Illinois Center for Advanced Studies of the Universe and Department of Physics, University of Illinois at Urbana-Champaign, Urbana, IL 61801, USA}

\begin{abstract}
We show that a rotating axion field that makes a transition from a matter-like equation of state to a kination-like equation of state around the epoch of recombination can significantly ameliorate the Hubble tension, i.e., the discrepancy between the determinations of the present-day expansion rate $H_0$ from observations of the cosmic microwave background on one hand and Type Ia supernovae on the other. 
We consider a specific, UV-complete model of such a rotating axion and find that it can relax the Hubble tension without exacerbating tensions in determinations of other cosmological parameters, in particular the amplitude of matter fluctuations $S_8$. 
We subsequently demonstrate how this rotating axion model can also generate the baryon asymmetry of our universe, by introducing a coupling of the axion field to right-handed neutrinos. This baryogenesis model predicts heavy neutral leptons that are most naturally within reach of future lepton colliders, but in finely-tuned regions of parameter space may also be accessible at the high-luminosity LHC and the beam dump experiment SHiP. 
\end{abstract}

\maketitle

\preprint{PITT-PACC-2402} 

\section{Introduction}
A rotating axion field in the early universe has proven to be a powerful tool in addressing the cosmological deficits of the Standard Model (SM).  
Rotating axions provide a class of mechanisms for baryogenesis, called axiogenesis~\cite{Chiba:2003vp,Takahashi:2003db,Kamada:2019uxp,Co:2019wyp,Co:2020xlh,Co:2020jtv,Harigaya:2021txz,Chakraborty:2021fkp,Kawamura:2021xpu,Co:2021qgl,Barnes:2022ren,Co:2022kul,Badziak:2023fsc,Berbig:2023uzs,Chao:2023ojl,Chun:2023eqc,Barnes:2024jap,Wada:2024cbe,Datta:2024xhg}, broaden the parameter space for axion dark matter via the kinetic misalignment mechanism~\cite{Co:2019jts,Co:2020dya,Co:2021rhi,Eroncel:2022vjg}, generate cosmic perturbations~\cite{Co:2022qpr}, and open new windows onto the early universe through gravitational waves \cite{Co:2021lkc,Gouttenoire:2021wzu,Co:2021rhi,Madge:2021abk,Harigaya:2023mhl,Harigaya:2023pmw}.  
In these models, a coherent axion field that is initially displaced from the origin of field space has interactions with a radiation bath that enable it to efficiently damp its initial radial oscillations while retaining sizeable rotational energy.
These interactions can also mediate the (partial) transfer of the axion's initial angular momentum to a particle number asymmetry in the radiation bath, enabling a range of natural baryogenesis scenarios.
The axion's ensuing cosmological evolution consists of coherent field rotations that transition from a matter-like ($\rho\propto 1/a^3$, with $a$ the scale factor of the universe) to a kination-like ($\rho\propto 1/a^6$) equation of state, with the details of the transition determined by the potential within a given model. We refer to this axion evolution as a whole as {\it axion kination}.

Another outstanding cosmological puzzle is posed by the persistent tensions between high- and low-redshift determinations of the Hubble constant $H_0$. Most strikingly, the most recent determination of $H_0$ from Cepheid-calibrated Type-Ia supernovae \cite{Murakami:2023xuy} disagrees at the $5.7\sigma$ level from the values preferred by best-fit $\Lambda$CDM models to Planck data \cite{Planck:2018vyg}. This discrepancy is part of a broader pattern, (see, e.g., Refs.~\cite{Riess:2019qba, Shah:2021onj, Abdalla:2022yfr, Kamionkowski:2022pkx, Hu:2023jqc, Vagnozzi:2023nrq} for reviews), where indirect determinations of $H_0$ from $\Lambda$CDM fits to cosmological datasets consistently prefer significantly lower values of $H_0$ than do local distance-ladder measurements.

In this paper, we construct a minimal model of axion baryogenesis that allows the rotating axion to make the transition to kination as late as the era of cosmic microwave background (CMB) formation while still successfully generating the baryon asymmetry of the universe.  We find that the Hubble tension can be substantially mitigated when a rotating axion field initially contributing $\sim 1\%$ of the matter density transitions from matter to kination near recombination.  
We also find that, unlike many other early-universe approaches to the Hubble tension, axion kination has the advantage of easing the Hubble tension without exacerbating tensions in other observables, as we discuss further below. 
  
In order for a rotating axion to contribute at the percent level to the matter density of the universe around recombination, its potential must be governed by relatively low mass scales. This is easiest to accomplish if the axion is not directly coupled to particles carrying SM charges, so that quantum corrections to the potential of the radial direction can be suppressed. We thus introduce SM-singlet right-handed neutrino states in order to transfer the axion's PQ charge to SM lepton number, which is subsequently reprocessed into baryon number by electroweak sphaleron processes.  We demonstrate here that successful baryogenesis together with a low axion matter-to-kination transition scale $a_c$ yields sharp predictions for the masses and interaction strengths of the right-handed neutrinos, which result in a motivated parameter space that is largely within future experimental reach.

The organization of this paper is as follows. We begin by discussing the cosmology of rotating axions in Sec.~\ref{sec:axion_cosmo} and the requirements for realizing a CMB-scale transition. 
In Sec.~\ref{sec:fit_to_data} we perform a fit to CMB, large-scale structure, and supernova data and quantify the degree to which axion kination can address the Hubble tension.  In Sec.~\ref{sec:baryons} we construct a minimal model of baryogenesis using CMB-scale axion kination, and discuss the terrestrial signatures that it predicts. 
Our conclusions are in Sec.~\ref{sec:discussion}.
Further details about the cosmology of the model, the evolution of axion perturbations, and the axion equation of state are discussed in Appendices~\ref{sec:two-field}-\ref{sec:transfer_rate}.

%%%%%%%%%%%%%%%%%%%%%%%%%%%%%%%%%
\section{Axion kination at the CMB epoch}
\label{sec:axion_cosmo}
%%%%%%%%%%%%%%%%%%%%%%%%%%%%%%%%%

The axion of interest to us is a Nambu-Goldstone boson that results from the spontaneous breaking of a global $U(1)$ symmetry, which we call a Peccei-Quinn (PQ) symmetry.
Explicit breaking of this PQ symmetry in the early universe can induce rotations in the angular, i.e., axion, direction of the field space~\cite{Affleck:1984fy}. This rotational motion can have a major impact on the subsequent cosmological evolution of the axion field.
In particular, we consider the axion cosmologies developed in Refs.~\cite{Co:2019wyp,Co:2020jtv,Co:2021lkc,Kawamura:2021xpu,Gouttenoire:2021jhk}, where a complex field $P=1/\sqrt{2} \, r e^{i \theta/f}$ initially rotates at a radial displacement away from the minimum of the potential.
While the axion rotation redshifts toward the radial minimum, the energy density of the axion field $\rho_{\rm rot}$ redshifts as matter.  Once the radial mode reaches the minimum of its potential, the rotations continue and the energy density now redshifts as kination.%
\footnote{
If the axion is massive,
the kination phase will ultimately end when the angular velocity is sufficiently close to the axion mass.
In this work, we assume that the axion is either massless or sufficiently light that $\rho_{\rm rot}$ redshifts as kination well past recombination.
}

We now describe the evolution of the rotation in more detail.
The complex field $P$ is assumed to take a large initial field value after inflation, as in the Affleck-Dine mechanism~\cite{Affleck:1984fy}.  At large field values, higher-dimensional operators in $P$ can be important for determining the field evolution. We consider higher-dimensional operators that violate the PQ symmetry, thereby providing a kick to the angular direction and initiating the rotations of $P$.

The initial axion rotation is generically elliptical, i.e.,  a superposition of circular rotations and radial oscillations. Interactions of $P$ with a thermal bath at temperature $T$ allow the coherent axion field to reach thermal equilibrium, after which the radial oscillation mode is dissipated. The circular rotation mode remains almost intact as long as the charge density stored in the axion field $n_{\mathrm{PQ}} = r^2 \dot\theta$ is larger than $m_P T^2$, where $m_P$ is the mass of the radial mode. This is because the free energy is minimized when most of the $U(1)$ charge is carried by the scalar field's coherent rotation rather than by particles in the bath~\cite{Co:2019wyp,Domcke:2022wpb} (see also~\cite{Laine:1998rg}).
Thus, after thermalization, the rotation becomes circular. 

The PQ charge density of the rotation decreases through cosmic expansion. As long as the radius of the circular rotation is larger than the value of the radial mode at the minimum of the potential $\sim v_{\rm PQ}$, the radius of rotation then decreases with the expansion of the universe to ensure charge conservation. If the potential of $P$ is nearly quadratic, the energy density of the rotation decreases as non-relativistic matter, $\rho_{\rm rot} \propto a^{-3}$. The radius eventually reaches the global minimum of the potential, after which the radius is fixed while the angular velocity decreases in proportion to $a^{-3}$. The energy density of the rotation, which is dominantly the kinetic energy, then decreases in proportion to $a^{-6}$. 

The time-dependence of the axion evolution depends in detail on the scalar field potential in a given model. We consider here the ``two-field'' model studied in~\cite{Co:2021lkc}. The circular motion of the axion is described in this model by the effective Lagrangian
\begin{align}
\label{eq:L}
    {\cal L} \simeq & 
    \frac{1}{2} F^2 \left(\partial_\mu \theta \partial^\mu\theta\right)  \\
    &- \frac{1}{4} m_P^2 F^2 \left(  \left(1+r_P^2\right) + \left( 1-r_P^2 \right) \sqrt{1 - \left( \frac{2 v_{\rm PQ}}{F} \right)^4 }   \right), \nonumber 
\end{align}
where $F\equiv \sqrt{2} r (1+v_{\rm PQ}^4/r^4)^{1/2}$ is the effective radial mode, $\theta$ is the angular mode, $m_P$ is the mass of the radial mode, $v_{\rm PQ}$ is the $U(1)$ symmetry breaking scale, and $r_P >1$ is the ratio of the masses of the ``two fields" in the UV model. During the rotation, the kinetic terms for $F$ are negligible. This model has four free parameters: $v_{\rm PQ}$, $r_P$, $m_P$, and the conserved PQ charge $n_{\rm PQ}$ stored in the rotating axion field, which is determined by the initial conditions.  The PQ charge controls the abundance of the axion field, while the parameters $v_{\rm PQ}$ and $r_P$ determine the value of $F$ at the minimum of the potential. 
See Appendix~\ref{sec:two-field} for a full discussion of the UV completion of this effective Lagrangian. The UV completion is supersymmetric, which is important for ensuring the scalar potential considered in this model remains a reliable description of the system over a wide range of scales.

In order to generate the observed baryon asymmetry from the axion rotation, the angular velocity $\dot\theta$ needs to be sufficiently large. As we will see in Sec.~\ref{sec:baryons}, the angular velocity that is observationally preferred for CMB-scale axion kination is too small to account for the baryon asymmetry of our universe within a quadratic potential.  We will therefore add a quartic term to the potential in order to explain the observed baryon asymmetry, as we discuss below in Sec.~\ref{sec:baryons}. The effects of this additional contribution to the potential are important at high temperatures when the SM baryon asymmetry is generated, but are negligible by the CMB epoch.  Thus, Eq.~\eqref{eq:L} remains the appropriate action governing  the effect of the rotating axion on the CMB power spectrum.

\begin{figure}
\includegraphics[width=\columnwidth]{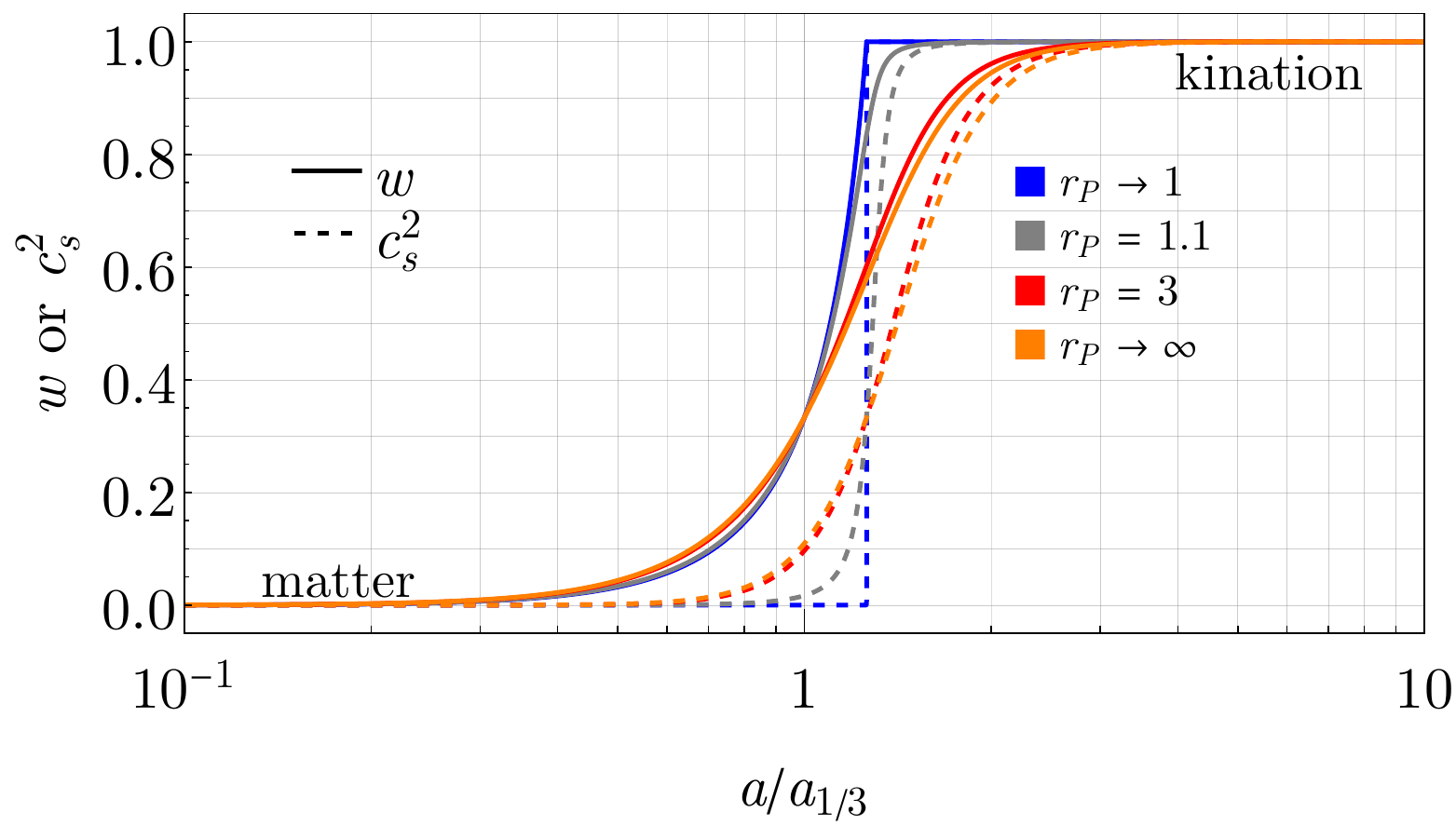}%  
\caption{The equation of state parameter $w$ (solid curves) and the adiabatic sound speed-squared $c_s^2$ (dashed curves) of the rotating axion as a function of the scale factor $a$, where $a = a_{1/3}$ when $w$ is equal to that of radiation $w = 1/3$. Here $r_P>1$ parameterizes the shape of the potential.
}
\label{fig:w_cs}
\end{figure}

Phenomenologically, this model's impact on cosmological observables can be characterized by three parameters: the overall abundance of the axion $f_{\mathrm{kin}}(a_c)$, the scale factor $a_c$ where the axion transitions from matter to kination, and $r_P$, which controls the equation of state. 
The fraction $f_{\mathrm{kin}}(a) \equiv \frac{\rho_{\rm rot}(a)}{\rho_{m}(a)+ \rho_{r}(a) + \rho_{\rm rot}(a)}$ refers to the ratio of the energy density in the rotating axion field $\rho_{\rm rot}$ to
the sum of matter $(\rho_{m})$, radiation $(\rho_{r})$, and rotation
energy densities. The scale factor $a_{c}$ denotes the scale factor when $f_{\mathrm{kin}}(a)$ is maximized, and controls the timing of the matter-kination transition.
Meanwhile the shape parameter $r_P$ determines how rapidly the rotating field transitions from a matter-like ($w = 0$) to a kination-like ($w = 1$) equation of state, as shown in 
the solid curves in Fig.~\ref{fig:w_cs}, where different colors represent different values of $r_P$ as indicated. The corresponding adiabatic sound speed-squared $c_s^2$ is shown by the dashed curves.
As seen in Fig.~\ref{fig:w_cs}, while $r_P$ can vary from unity to infinity, the corresponding variation of the curve $w(a)$ is limited. 
A combination of all four model parameters determines $a_c$, which together with $n_{\rm PQ}$ then determines $f_{\rm kin}(a_c)$.
The homogeneous cosmology of this two-field model, including the derivation of the function $w(a)$, and the associated Boltzmann equations describing the evolution of axion perturbations are described in detail in Appendices~\ref{sec:Pert_eqs} and~\ref{sec:eos}.

%%%%%%%%%%%%%%%%%%%%%%%%%%%%%%%%%%
%%%%%%%%%%%%%%%%%%%%%%%%%%%%%%%%%%

\begin{figure*}
\includegraphics[width=0.496\linewidth]{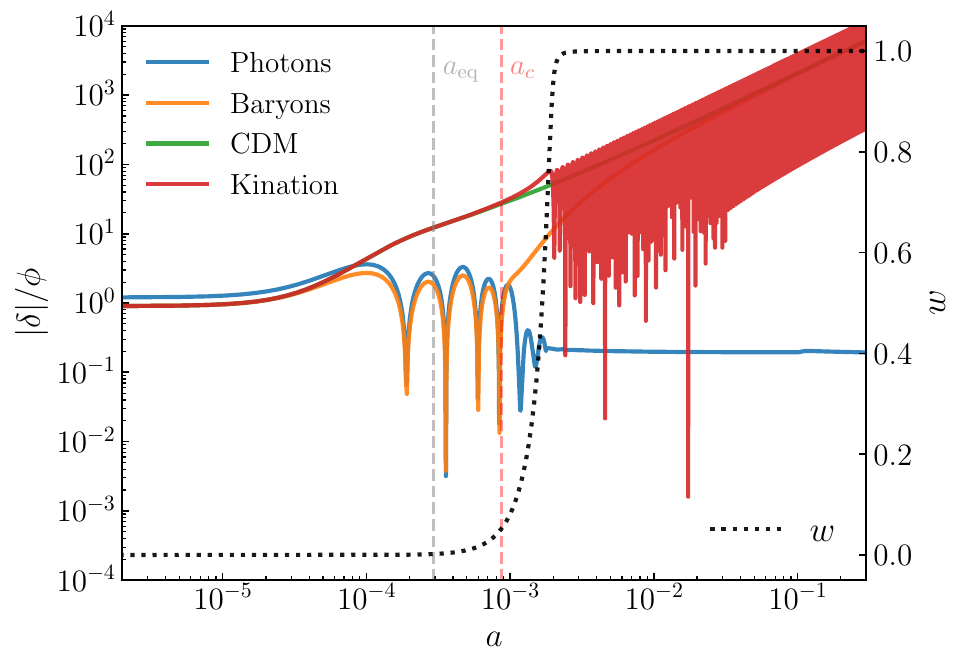}
\includegraphics[width=0.496\linewidth]{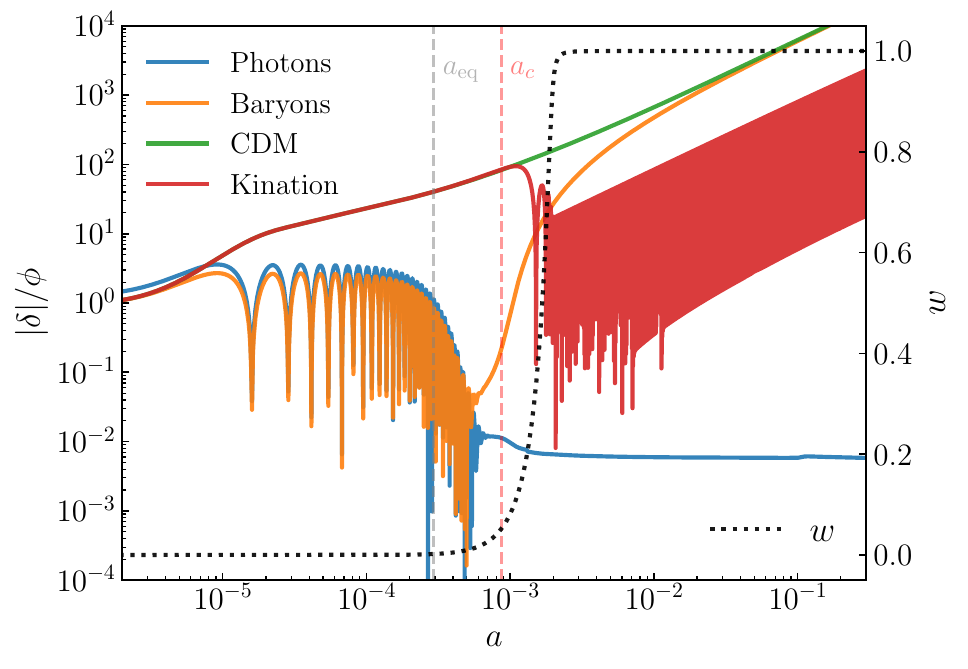}
\caption{\label{fig:perturbations} Evolution of the synchronous gauge density perturbation $\delta$ with scale factor for two different modes, $k=0.1/$Mpc (left) and $k=1/$Mpc (right), in units of the primordial curvature perturbation $\phi$. Cosmic parameters are chosen according to the best-fit values when fitting to the data set $\mathcal{DH}$, given in Table.~\ref{tab:summary}. We show the kinaton perturbations in red, along with photons (blue), baryons (orange), and cold DM (green). Vertical dashed lines indicate $a_{\rm eq}$ and $a_c$. We additionally show the kinaton equation of state in the dotted black line. Here we fix $r_P = 1.1$.}
\label{fig:perts}
\end{figure*}

We show the evolution of density perturbations in the two-field model for two values of comoving wavenumber ($k=0.1/$Mpc and $k=1/$Mpc) in Fig.~\ref{fig:perts}. Here we choose a representative $r_P=1.1$, and choose $f_{\rm kin}(a_c)$ and $a_c$ according to the best-fit values of cosmic parameters given in Table~\ref{tab:summary} below. The axion equation of state is superimposed for reference (dotted line). The kination perturbations follow those of cold dark matter (CDM) while $w\approx 0$, and begin to oscillate rapidly when the variation of $w$ with scale factor becomes appreciable.  Note that for the indicated value of $a_c$ the onset of kination oscillations occurs after the end of the baryon drag epoch. 

%%%%%%%%%%%%%%%%%%%%%%%%%%%%%%%%%%
%%%%%%%%%%%%%%%%%%%%%%%%%%%%%%%%%%%%

As we will see in Sec.~\ref{sec:fit_to_data}, an axion transitioning from matter to kination during the CMB epoch yields a good fit to observations when the energy stored in the rotating axion is $\mathcal{O}(1\%)$ of the total energy around the time of recombination, more precisely $f_{\rm kin} \simeq 0.01$ at $T = T_c \simeq 0.3$~eV. For such a low-scale transition, we need $\left. m_P^2 v_{\rm PQ}^2 \right|_{T_c} \simeq \left( 0.13~{\rm eV} \right)^4$. The perturbativity of the complex-field potential requires $m_P < v_{\rm PQ}$, giving
\begin{equation}
\label{eq:mP_max}
    m_P \lesssim 0.13 ~{\rm eV} \,.
\end{equation}
With such a small energy scale, such an axion is most readily thermalized within a dark sector that is sequestered from both the SM and the supersymmetry-breaking sector, so that the supersymmetry-breaking mass term $m_P$ is naturally suppressed relative to SM scales.   The thermalization required to dissipate the radial oscillations must then proceed via a decoupled dark sector bath.
We require such thermalization to occur before coherent radial axion oscillations dominate the dark sector, in order to avoid scenarios with an unacceptably large dark radiation energy density. 
At the time of thermalization, the energy density of the rotation is of the same order as that of the dark radiation, but the rotating axion will come to dominate over the dark radiation bath during the period where it redshifts as matter. Thus, by recombination the contribution of the dark radiation to the energy density of the universe is generically negligible. 
We also comment that the requirement that $m_P^2 v_{\rm PQ}^2\simeq \left( 0.13~{\rm eV} \right)^4 $ picks out an energy scale substantially below the QCD scale. Thus, this rotating axion cannot be identified with the QCD axion, or, in other words, it cannot solve the strong CP problem.

%%%%%%%%%%%%%%%%%%%%%%%%%%%%%%%%%%%%%
\section{Fit to Data}
\label{sec:fit_to_data}
%%%%%%%%%%%%%%%%%%%%%%%%%%%%%%%%%%%%%

In order to establish the impact of CMB-scale kination on cosmological observables, 
we implement the axion kination cosmology and the associated perturbation equations in CLASS~\cite{Blas:2011rf}. 
In addition to studying how the rotating axion affects the Hubble parameter $H_0$, we also consider its impact on $S_8\equiv \sigma_8\sqrt{\Omega_m/0.3}$, where $\sigma_8$ is the root-mean-square amplitude of matter fluctuations at the scale $8 \,\mathrm{Mpc}/h$ and $\Omega_m$ is the fraction of the critical density constituted by matter. Low-redshift and high-redshift observations of $S_8$ also exhibit a persistent discrepancy, albeit less significant than discrepancies in measurements of $H_0$. A wide range of weak gravitational lensing measurements and galaxy cluster surveys, reviewed in \cite{Abdalla:2022yfr},
favor values of $S_8$ that are $\sim 2$-$3\sigma$ lower than the values preferred by Planck. 
Measurements of $S_8$ that have appeared after this review have, however, tended toward somewhat higher values of $S_8$.  
Both weak lensing measurements \cite{Kilo-DegreeSurvey:2023gfr} and determinations using quasars \cite{Alonso:2023guh} exhibit reduced tension but still prefer values below Planck's best fit, at the 1.5$\sigma$ level. 

Measurements of $S_8$ are particularly of interest in the context of the Hubble tension, since models that address the discrepancy in $H_0$ by altering the early universe sound horizon typically predict {\it higher} values of $S_8$, exacerbating the tension \cite{Knox:2019rjx, Jedamzik:2020zmd}.  In order to avoid conflicts with observations of structure formation, models that aim to address the Hubble tension are thus generally required to invoke multiple ingredients active at different cosmological epochs, e.g.~\cite{Clark:2021hlo,Ye:2021iwa,Buen-Abad:2022kgf,Wang:2022bmk,Bansal:2022qbi,Cruz:2023lmn, Khalife:2023qbu}, though see \cite{Schoneberg:2022grr, Joseph:2022jsf, Buen-Abad:2023uva, Allali:2023zbi, Schoneberg:2023rnx, Aloni:2021eaq}.

We perform a combined fit to several cosmological datasets.  We follow the procedure laid out in~\cite{Schoneberg:2021qvd} in order to facilitate comparison with other models, although we use the updated SH0ES result~\cite{Riess:2021jrx}. 
\begin{itemize}
\item Our baseline dataset $\mathcal{D}$ consists of (i) low- and high-$\ell$ temperature and polarization power spectra and lensing from Planck \cite{Planck:2019nip}; (ii) the baryon acoustic oscillation (BAO) measurements from BOSS DR12 \cite{BOSS:2016wmc}, MGS \cite{Ross:2014qpa}, and 6dFGS \cite{Beutler:2011hx}; and (iii) the type Ia supernovae apparent magnitudes from Pantheon \cite{Pan-STARRS1:2017jku}. 
\item Combining the previous measurements with the SH0ES measurement of the absolute supernova luminosity calibration that translates to $H_0=73.04 \pm 1.04 \, \mathrm{km/s/Mpc}$ \cite{Riess:2021jrx} gives a dataset that we denote as $\mathcal{DH}$.
\item Finally, we impose weak lensing and galaxy clustering measurements of $S_{8}=0.790^{+0.018}_{-0.014}$ resulting from the recent joint cosmic shear analysis of the Dark Energy Survey (DES Y3) and the Kilo-Degree Survey (KiDS-1000) \cite{Kilo-DegreeSurvey:2023gfr}. We denote the combination with the previous datasets $\mathcal{DHS}$. 
\end{itemize}

To find the best fit values of our cosmological parameters, we perform a Markov chain Monte Carlo (MCMC) analysis using MontePython \cite{Brinckmann:2018cvx} and consider chains to be converged if the Gelman-Rubin criterion \cite{Gelman:1992zz} $|R-1|\leq 0.01$ is met. We adopt flat priors for our cosmological parameters: the $\Lambda$CDM parameter set $\{\omega_\mathrm{b}, \omega_\mathrm{c}, H_0, \log_{10}(10^{10} A_\mathrm{s}), n_\mathrm{s}, \tau\}$, to which we add the kinaton parameters $\{ f_{\rm kin} (a_c), \log_{10}(a_c), r_P\}$. Here, as usual, we define $\omega_{\mathrm{b},\mathrm{c}} \equiv \Omega_{\mathrm{b},\mathrm{c}} h^2$ and require $f_{\mathrm{kin}} (a_c) < 1$.  
For the minimization of the $\chi^2$ values, we employ simulated annealing (a method outlined in \cite{Schoneberg:2021qvd}).
Following best practices \cite{Benevento:2020fev,Camarena:2021jlr,Efstathiou:2021ocp}, we incorporate SH0ES results as a prior on the absolute supernova luminosity calibration.

\begin{figure*}[ht]
\includegraphics[width=2.0\columnwidth]{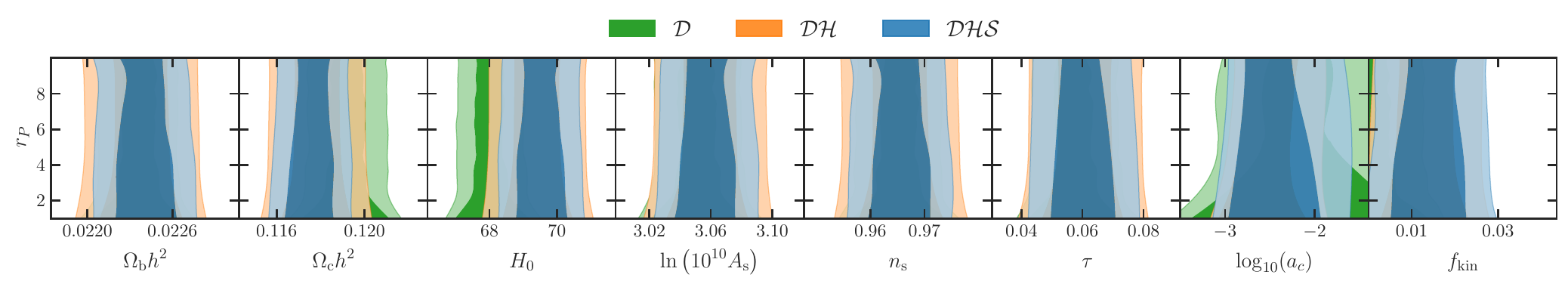}% 
\caption{\label{fig:Posteriors_rp} Contours of the 2D posterior distributions for $r_P$ versus the other eight model parameters resulting from fits to the $\mathcal{D}$ (green), $\mathcal{DH}$ (orange) and $\mathcal{DHS}$ (blue) datasets. Allowing $r_p$ to vary makes no significant difference in posterior distributions for other model parameters.}
\end{figure*}

We find that predictions of our two-field model strongly depend on $a_c$ and $f_{\rm kin}$, but are insensitive to $r_P$. This can be seen in Fig.~\ref{fig:Posteriors_rp}, which shows how the 2D reconstructed posteriors for the other eight model parameters are insensitive to $r_P$ as it varies over the range $1<r_P <10$, which covers the physical variation of the equation-of-state curve $w(a)$. Thus, for our main results we fix a reference value of $r_P=1.1$ and treat the other two model parameters ($a_{c}, f_{\rm kin}$) as free.  

%%%%%%%%%%%%%%%%%%%%%%%%%%%%%%%
%%%%%%%%%%%%%%%%%%%%%%%%%%%%%%%

Our results from this eight-parameter fit are tabulated in Table \ref{tab:summary}.
In addition to the overall $\Delta\chi^2$ relative to $\Lambda$CDM, we present two statistical tests adopted from \cite{Schoneberg:2021qvd}. First, the Akaike Information Criterion (AIC),
\begin{equation}
    \label{eq:aic}
\Delta \mathrm{AIC} =  \chi_{\mathrm{min, \,kin}}^{2} - \chi_{\mathrm{min,\,\Lambda CDM}}^{2} + 2 (N_{\rm kin} - N_{\mathrm{\Lambda CDM}}),
\end{equation}
compares the two models' fits with a penalty for the number of parameters, indicated by $N_{\mathrm{kin}}$ ($N_{\Lambda\mathrm{CDM}}$) in our kination 
($\Lambda$CDM) model.
Second, the $Q_{\mathrm{DMAP}}$ tension,
\begin{equation}
\label{eq:qdmap}
    {\rm Q}_{\rm DMAP}=\sqrt{\chi_{\mathcal{D+\mathrm{SH0ES}}}^{2}-\chi_{\mathcal{D}}^{2}},
\end{equation}
quantifies the difference in the best-fit $\chi^{2}$ for fits with and without $\rm{SH0ES}$. We find that the presence of axion kination improves the best-fit $\chi^{2}$ to the $\mathcal{DH}$ dataset by $\Delta\chi^2 = -8.04$ ($\Delta \mathrm{AIC} = -4.04$) compared to $\Lambda$CDM.  It also presents a meaningful but modest improvement in the $Q_{\rm DMAP}$ statistic, decreasing the tension from $5.7 \sigma$ to $4.4 \sigma$.
This improvement in $\Delta \mathrm {AIC}$ indicates a ``weak'' preference for the axion kination model over $\Lambda$CDM by the metric proposed in \cite{Schoneberg:2021qvd}.  The  preference for axion kination over $\Lambda$CDM is largely driven by the SH0ES measurement, as indicated by the positive value of $\Delta \mathrm{AIC}$ we find from the fit to $\mathcal{D}$ alone.  
While this preference is not strong, we nonetheless find it notable: CMB-scale axion kination is a novel phenomenological ingredient that mitigates the Hubble tension without inducing large shifts in other cosmic parameters, notably $S_8$, as well as providing a viable avenue for baryogenesis.

 %%%%%%%%%%%%%%%%%%%%%%%%%
\begin{table}
\begin{tabular}{|c|ccc|ccc|} 
 \hline 
Model &  \multicolumn{3}{c|}{$\Lambda$CDM} & \multicolumn{3}{c|}{Axion kination} \\ \hline \hline
Dataset & \multicolumn{1}{c|}{$\mathcal{D}$} & \multicolumn{1}{c|}{$\mathcal{DH}$} & \multicolumn{1}{c|}{$\mathcal{DHS}$} & \multicolumn{1}{c|}{$\mathcal{D}$} & \multicolumn{1}{c|}{$\mathcal{DH}$} & $\mathcal{DHS}$ \\ \hline 
$\Omega_{\mathrm{b} } h^{2}$  & \multicolumn{1}{c|}{$0.0224$}  & \multicolumn{1}{c|}{$0.0226$} & \multicolumn{1}{c|}{$0.0226$} & \multicolumn{1}{c|}{$0.0223$}  & \multicolumn{1}{c|}{$0.0223$} & 0.0224  \\
$\Omega_{\mathrm{c} } h^{2}$  & \multicolumn{1}{c|}{$0.1192$} & \multicolumn{1}{c|}{$0.1180$} & \multicolumn{1}{c|}{$0.1178$} & \multicolumn{1}{c|}{$0.1192$} & \multicolumn{1}{c|}{$0.1180$} & 0.1180 \\
$H_0$ &\multicolumn{1}{c|}{$67.75$}  & \multicolumn{1}{c|}{$68.42$} & \multicolumn{1}{c|}{$68.47$} & \multicolumn{1}{c|}{$68.23$}  & \multicolumn{1}{c|}{$69.70$} & 69.71  \\
$\ln (10^{10}A_{s })$  &\multicolumn{1}{c|}{$3.049$}  & \multicolumn{1}{c|}{$3.0542$} & \multicolumn{1}{c|}{$3.0540$} & \multicolumn{1}{c|}{$3.051$}  & \multicolumn{1}{c|}{$3.060$} & 3.054  \\
$n_{s}$  &\multicolumn{1}{c|}{$0.9680$}  & \multicolumn{1}{c|}{$0.9712$} & \multicolumn{1}{c|}{$0.9715$} & \multicolumn{1}{c|}{$0.9712$} & \multicolumn{1}{c|}{$0.9682$} & 0.9690\\
$\tau_{\mathrm{reio}}$  &\multicolumn{1}{c|}{$0.0590$}  & \multicolumn{1}{c|}{$0.0601$} & \multicolumn{1}{c|}{$0.0604$} & \multicolumn{1}{c|}{$0.0570$}  & \multicolumn{1}{c|}{$0.0609$} & 0.0588 \\
$\log_{10} (a_{c})$ & \multicolumn{1}{c|}{-} & \multicolumn{1}{c|}{-} & \multicolumn{1}{c|}{-} & \multicolumn{1}{c|}{$-3.091$} & \multicolumn{1}{c|}{$-3.058$} & $-3.070$ \\
$f_{\mathrm{kin} }$ & \multicolumn{1}{c|}{-} & \multicolumn{1}{c|}{-} & \multicolumn{1}{c|}{-} & \multicolumn{1}{c|}{$0.005$} & \multicolumn{1}{c|}{$0.0127$} & 0.0118  \\ \hline
$\sigma_8$ &\multicolumn{1}{c|}{$0.8120$} & \multicolumn{1}{c|}{$0.8085$} & \multicolumn{1}{c|}{$0.8086$} & \multicolumn{1}{c|}{$0.8157$} & \multicolumn{1}{c|}{$0.8226$} & 0.8194 \\
$\Omega_{m }$ &\multicolumn{1}{c|}{$0.290$}  & \multicolumn{1}{c|}{$0.3016$} & \multicolumn{1}{c|}{$0.291$} & \multicolumn{1}{c|}{$0.3055$}  & \multicolumn{1}{c|}{$0.2915$} & 0.2895 \\
$S_{8 }$ &\multicolumn{1}{c|}{$0.825$}  & \multicolumn{1}{c|}{$0.811$} & \multicolumn{1}{c|}{$0.796$} & \multicolumn{1}{c|}{$0.823$}  & \multicolumn{1}{c|}{$0.809$} & 0.805  \\
\hline
\hline
$\Delta \chi^2_{\mathrm{tot}}$ &\multicolumn{1}{c|}{$0$} & \multicolumn{1}{c|}{$0$} & \multicolumn{1}{c|}{$0$} & \multicolumn{1}{c|}{$-0.38$} & \multicolumn{1}{c|}{$-8.04$} & $-6.96$ \\
$\Delta$AIC & \multicolumn{1}{c|}{$-$} & \multicolumn{1}{c|}{$-$} & \multicolumn{1}{c|}{$-$} & \multicolumn{1}{c|}{$3.62$}  & \multicolumn{1}{c|}{$-4.04$} & $-2.96$  \\
$Q_{\rm DMAP}$ & \multicolumn{1}{c|}{$-$} & \multicolumn{1}{c|}{$5.7$} & \multicolumn{1}{c|}{$-$} & \multicolumn{1}{c|}{$-$} & \multicolumn{1}{c|}{$4.4$} & $-$  \\
\hline 
 \end{tabular}
 \caption{\label{Parameters} Parameters of the best-fit model for $\Lambda$CDM and axion kination resulting from fits to the datasets $\mathcal{D}$ and $\mathcal{DH}$. The different criteria to measure model success---Akaike Information Criterion (AIC) and difference of the maximum {\it a posteriori} $Q_{\rm DMAP}$---are defined in Eqs.~(\ref{eq:aic}) and~(\ref{eq:qdmap}).
 }
 \label{tab:summary}
\end{table}
%%%%%%%%%%%%%%%%%%%%%%%%%

%%%%%%%%%%%%%%%%%%%%%%%%%%%%%%
 \begin{figure*}
\includegraphics[width=0.95\linewidth]{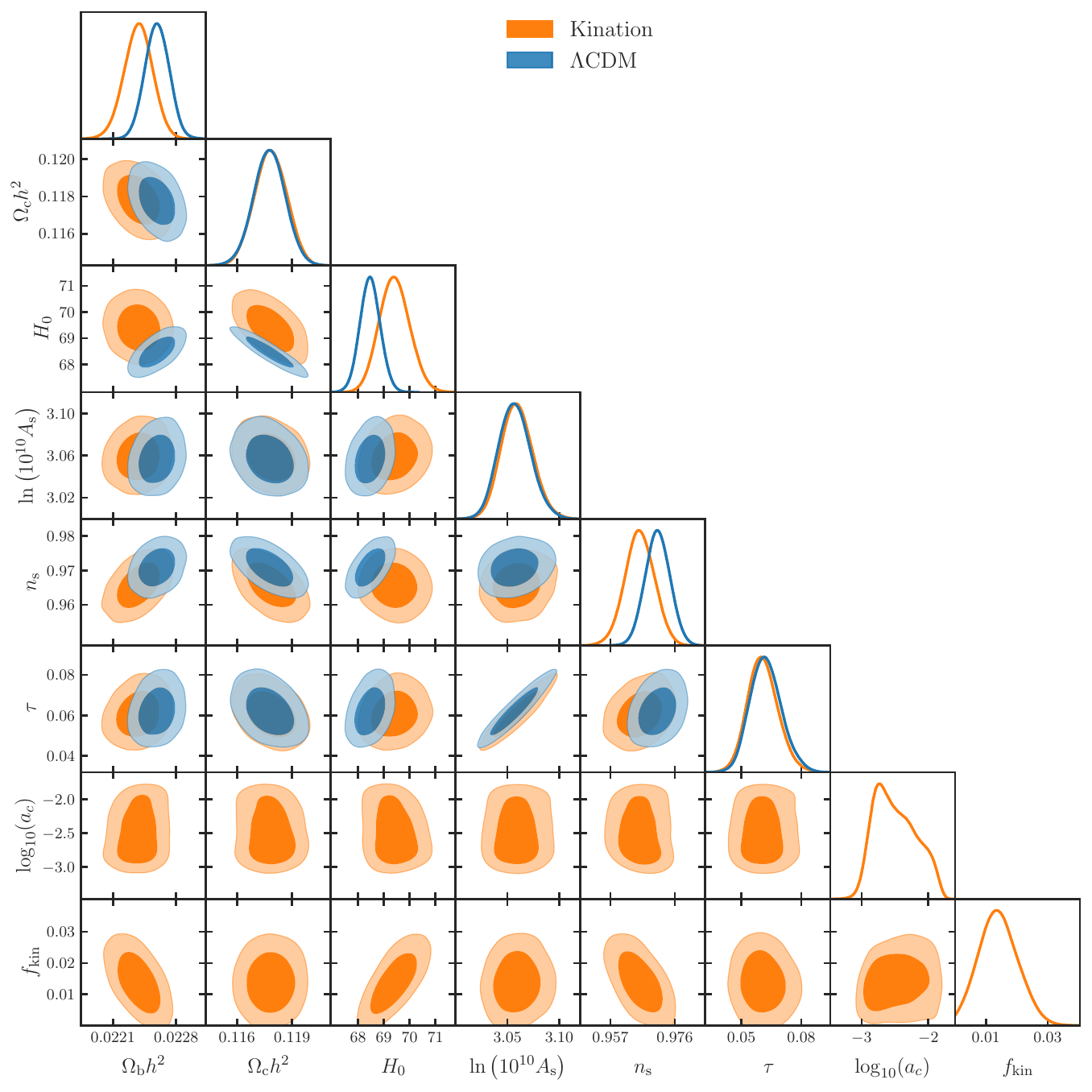}% 
\caption{\label{fig:Posteriors}Contours of the 2D posterior distributions for $\Lambda$CDM (blue) and kination (orange) resulting from fits to the dataset~$\mathcal{DH}$.}
\end{figure*}
%%%%%%%%%%%%%%%%%%%%%%%%%%%%%%%

Posterior distributions for several quantities of interest are shown in Fig.~\ref{fig:Posteriors}. We find that the data prefer a small but nonzero fraction of energy in kination, with best-fit $f_{\rm kin} =1.3\%$, that transitions from matter to kination at a scale factor $a_c$ {\it after} the baryon drag epoch.  Thus, in the best-fit cosmology, axion kination serves as a sub-component of CDM that subsequently redshifts away shortly after baryon-photon decoupling is complete.  This extra contribution to the effective CDM abundance in the early universe yields a larger sound horizon given the same late-time CDM density, and correspondingly, a larger Hubble constant. The post-recombination shift from matter to kination gives rise to a novel contribution to the early integrated Sachs-Wolfe effect that, for the preferred values of $a_c$, primarily serves to adjust the height of the first acoustic peak. 

The addition of a cold matter-like component through much of recombination, without any additional contributions to the radiation density, alters the timing of matter-radiation equality and gives rise to less radiation damping at high $\ell$. This drives the notable shift to lower values of $n_\mathrm{s}$, relative to the $\Lambda$CDM fit to $\mathcal{DH}$, in order to provide less primordial power on small scales. 
This effect is what allows the axion kination model to raise the Hubble constant without increasing $n_\mathrm{s}$ to levels that induce tension with other cosmological datasets.
Meanwhile, the shift in $\omega_\mathrm{b}$ both compensates for $n_\mathrm{s}$-induced changes to the second acoustic peak and helps to adjust the damping tail by shifting the damping scale to larger $\ell$.  The small shift in $\tau$, on the other hand, increases damping, and helps to fit the amplitudes of peaks in the power spectrum in combination with the changes in $\omega_\mathrm{b}$ and $n_\mathrm{s}$.

\begin{figure}[b]
\includegraphics[width=0.98\columnwidth]{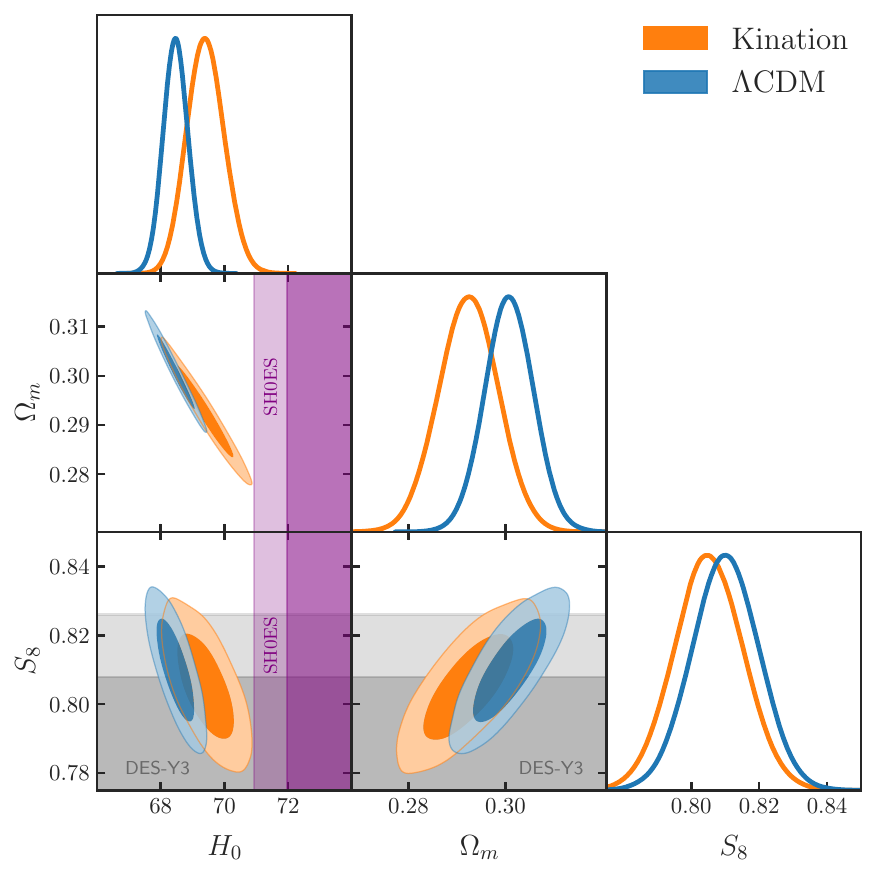}
\caption{Contours of the posterior distributions for $H_0$, $\Omega_m$ and $S_8$ for $\Lambda$CDM (blue) and with the addition of kination (orange), for a fit to the dataset $\mathcal{DH}$. Shaded purple and grey bands represent respectively the SH$0$ES result for $H_0$ \cite{Riess:2021jrx} and the DES{-}Y3 measurement of $S_{8}=0.790^{+0.018}_{-0.014}$) \cite{Kilo-DegreeSurvey:2023gfr}.}
\label{fig:H0S8}
\end{figure}

Adding kination to the fit allows us to increase $H_0$ while leaving $\omega_\mathrm{c}$ nearly fixed. Consequently,  
$\Omega_m$ decreases at the $\sim 1 \sigma$ level. 
This decrease in $\Omega_m$ allows us to mitigate the Hubble tension {\it without} further exacerbating the $S_8$ tension. We summarize the implications for $\Omega_m$, $S_8$ and $H_0$ in Fig.~\ref{fig:H0S8}, where we compare the predictions of axion kination to those of $\Lambda$CDM.

%%%%%%%%%%%%%%%%%%%%%%%%%%%%%%
%%%%%%%%%%%%%%%%%%%%%%%%%%%%%%

We close our discussion of the fit to cosmological data by placing CMB-scale axion kination in context with two other classes of models that aim to address the Hubble tension.  
Axion kination, with its unique equation of state, provides a previously unexplored phenomenological ingredient for addressing the Hubble tension: it increases the sound horizon in the early universe by adding a time-dependent component to the effective cold matter density during recombination.  This is somewhat similar to the mechanism invoked in decaying warm DM solutions \cite{Blinov:2020uvz, Holm:2022eqq,Nygaard:2023gel}.
However, unlike decaying warm DM models, axion kination results in a sudden change in the effective matter density, and does not need to invoke a separate dark radiation fluid.  Our results show axion kination also offers a substantially better fit than does decaying warm DM \cite{Holm:2022eqq}.

Axion kination involves a scalar field with a time-varying equation of state and is similar in that regard to models of early dark energy (EDE) \cite{Karwal:2016vyq, Poulin:2018dzj, Poulin:2023lkg}. While axion kination does not accommodate as large a value for $H_0$ as EDE can provide, the best-fit EDE models must also increase both the cold DM density $\Omega_\mathrm{c} h^2$ and the scale index $n_\mathrm{s}$ \cite{Hill:2020osr, Vagnozzi:2021gjh}. These increases significantly exacerbate the $S_8$ tension and may introduce new tensions with measurements of Ly-$\alpha$ absorption spectra~\cite{Goldstein:2023gnw}.  By contrast, the best-fit axion kination solution marginally decreases $S_8$ and prefers a value of $n_\mathrm{s}$ only slightly larger than the Planck result \cite{Planck:2018vyg}.  Thus, with the addition of a single new constituent in the early universe, following from a well-defined and UV-complete Lagrangian theory, we are able to significantly mitigate the Hubble tension without introducing further tension in measurements of $S_8$. We now turn to realizing baryogenesis within this scenario.

%%%%%%%%%%%%%%%%%%%%%%%%%%%%
\section{Baryogenesis from CMB-scale axion kination}
\label{sec:baryons}
%%%%%%%%%%%%%%%%%%%%%%%%%%%%

A CMB-scale axion model is dramatically different from the typical QCD axion or axion-like particle because the required decay constant $\sim v_{\rm PQ}$ is at most the eV scale.
This very small decay constant means that the axion must avoid having direct couplings to SM-charged particles in order to remain consistent with experimental bounds. Another issue for the axion kination cosmology is the need to protect the  potential of the radial direction against large quantum contributions from SM superpartners. 
We thus consider axion couplings to SM gauge singlets,
which makes an interaction through the right-handed neutrinos a particularly promising possibility. Related work in this direction for large axion decay constants can be found in Refs.~\cite{Co:2020jtv,Chakraborty:2021fkp,Kawamura:2021xpu,Barnes:2022ren,Berbig:2023uzs,Chao:2023ojl,Chun:2023eqc,Datta:2024xhg}. 

In what follows, we discuss a specific model as a proof of principle. We consider the model at temperatures above the electroweak scale, so that the complex field $P$ rather than the effective radial mode $F$ is the natural degree of freedom to consider.  We consider an inverse-seesaw-like coupling of $P$ with a right-handed neutrino $N$, so that the neutrino interactions in the Lagrangian are given by
\begin{equation}
\label{eq:L_transfer}
    \mathcal{L} = \frac{1}{2}\lambda P N^2 + m_N N \bar N + y LHN + {\rm h.c.} 
\end{equation}
Here the complex field interacts with $N$, while $N$ interacts with the Standard Model Higgs $H$ and left-handed lepton doublet $L$. These interactions allow the axion's PQ charge to be partially transferred to a SM lepton asymmetry via  $P N \leftrightarrow L^\dag H^*$ scattering. The transfer rate is given by 
\begin{equation}
\label{eq:rate}
    \Gamma_N \simeq \frac{3 \, \zeta(3)}{64 c_{B-L}\pi^3} \frac{\lambda^2 y^2 r^2}{T},
\end{equation}
where $r$ is the background value of the radial degree of freedom in $P$ and the coefficient $c_{B-L}$ is $1225/3084, 1265/1662$, or $145/132$ in the cases where $1,2,$ or $3$ generations of $N$ couple to $LH$, respectively. The derivation of the transfer rate is discussed in Appendix~\ref{sec:transfer_rate}. The lepton asymmetry then gets reprocessed into a baryon asymmetry by electroweak sphaleron processes. 
When the scattering rate of Eq.~(\ref{eq:rate}) is larger than the Hubble rate before the electroweak phase transition at $T_{\rm EW}$, the resulting baryon asymmetry is given by its equilibrium value,
\begin{equation}
\label{eq:YB}
    Y_B = \frac{c_B \dot\theta T^2}{s},
\end{equation}
with $s = 2\pi^2g_*T^3/45$ the SM entropy density and $c_B$ a coefficient governing the efficiency of charge transfer that can be derived from the detailed balance relations among SM chemical potentials. In the simple model of Eq.~\eqref{eq:L_transfer}, assuming equilibrium for strong and weak sphalerons and all SM Yukawa processes, we find $c_B = 21/257, 42/277,$ and $7/33$ for $1,2,$ and $3$ generations of $N$ in the thermal bath, respectively. In deriving these values, we also impose detailed balance relations from the conservation of hypercharge as well as the equilibrium of $L_iHN_i$ and $P^\ast N_i^\dag L_i H$ interactions for each generation of $N_i$ coupled to $L_i$ and, for other generations of $L_j$ not coupled to any $N$, the conservation of $B/3 - L_j$. 

The baryon asymmetry is fixed when the sphaleron processes go out of equilibrium, for which we use the value of $T_{\rm EW} = 130 ~{\rm GeV}$ predicted in the SM~\cite{DOnofrio:2014rug}. To explain the observed baryon asymmetry $Y_B \simeq 8.6 \times 10^{-11}$, the required angular velocity is
\begin{equation}
\label{eq:dottheta_EW}
    \dot\theta (T_{\rm EW}) \sim 0.5 ~{\rm keV} ~ c_B^{-1} 
    \left( \frac{T_{\rm EW}}{130 ~{\rm GeV}} \right),
\end{equation}
which is valid for $m_N < 130$ GeV, while we discuss the case with $m_N > 130$ GeV at the end of this section.
Since  $\dot\theta =\sqrt{(\partial V / \partial r)/r}$ but we require $m_P <$ eV, the quadratic potential of Eq.~(\ref{eq:L}) does not provide a large enough value of $\dot\theta$ to account for the baryon asymmetry. 
Therefore we add a quartic interaction, $\lambda_r r^4$, to the potential in Eq.~(\ref{eq:two-field}), so that $\dot\theta$ is enhanced at early times, especially at $T_{\rm EW}$, while the late-time behavior converges to the equation of state considered in the previous sections.

We now use Eq.~(\ref{eq:dottheta_EW}) together with the equations governing the homogeneous background evolution of the axion field (see Appendix~\ref{sec:Pert_eqs}) to determine the evolution of $\dot\theta(T)$ as well as $r(T)$. When the quartic term dominates, $\dot\theta(T) \propto T$. Then when the quadratic term dominates, $\dot\theta(T)$ approaches a constant value $m_P$.  The transition temperature $T_q$ between the quartic and quadratic regimes can be derived from $\dot\theta(T_{\rm EW}) (T_q/T_{\rm EW}) = m_P$ as
\begin{equation}
\label{eq:T_q}
    T_q \simeq 25 ~{\rm MeV} c_B \left( \frac{m_P}{0.1 ~{\rm eV}} \right).
\end{equation}
This transition temperature is significantly larger than the  eV-scale temperatures probed by the CMB, and thus the axion imprint on CMB anisotropies is well-described by the quadratic potential.
We can then derive the field evolution using the scalings $r \propto T^{3/2}$ and $r \propto T$ for $T < T_q$ and $T > T_q$, respectively. 

We now examine various constraints on the model and identify the viable parameter space. 
First, the consistency condition for the asymmetry transfer rate to reach equilibrium requires that $\Gamma_N \gtrsim H$ at $T_{\rm EW}$, resulting in a lower bound on the product of $\lambda y$ as
\begin{equation}
\label{eq:eq_con}
    \lambda y \gtrsim 3 \times 10^{-11}
    \left( \frac{c_{B-L}}{c_B} \right)^{\scalebox{1.01}{$\frac{1}{2}$} }
    \left( \frac{m_P}{0.1 ~{\rm eV}} \right)^{\scalebox{1.01}{$\frac{1}{2}$} }.
\end{equation}
Meanwhile, $N$ is assumed to be in thermal equilibrium at $T_{\rm EW}$ through the interaction $yLHN$, which requires
\begin{equation}
    y \gtrsim 4 \times 10^{-8}  .
\end{equation}

We also need to ensure that $N$ is in the bath at $T_{\rm EW}$ as assumed, which requires $\lambda r(T_{\rm EW}) < T_{\rm EW}$ and thus
\begin{equation}
    \lambda \lesssim  10^{-4} c_B^{-1/2} \left( \frac{m_P}{0.1 ~{\rm eV}} \right)^{\scalebox{1.01}{$\frac{1}{2}$} } .
\end{equation}

The radial mode receives a two-loop quantum correction to the mass of order $\Delta m_P^2 \sim \lambda^2 y^2 m_{\tilde L}^2 \ln^2(m_{\tilde L} /\Lambda) / (8 \pi^2)^2$ with $m_{\tilde L}$ the slepton mass. Specifically, $r$ receives a radiative mass correction from the sneutrino $\tilde N$, $d \, m_P^2 / d\ln\mu = \lambda^2 m_{\tilde N}^2/8\pi^2$, whose mass arises similarly from the loop correction of the slepton $d \, m_{\tilde N}^2 / d\ln\mu = y^2 m_{\tilde L}^2/4\pi^2$.
Requiring that $m_P$ remain sufficiently light then gives an upper bound on the product of $\lambda y$ as
\begin{equation}
\label{eq:quantum_correct}
    \lambda y \lesssim \frac{10^{-11}}{\sqrt{\delta_m}} 
    \left( \frac{m_P}{0.1 ~{\rm eV}} \right)
    \left(\frac{100 ~{\rm GeV}}{m_{\tilde L}} \right) 
    \left(\frac{6.9}{\ln(\Lambda/m_{\tilde L})} \right),
\end{equation}
where we assume $\Lambda = 10^5$ GeV, which is possible for low-scale gauge mediated supersymmetry breaking.%
\footnote{
The large log corrections can be avoided in Dirac gaugino scenarios~\cite{Fox:2002bu,Arvanitaki:2013yja,Csaki:2013fla,Alves:2015kia}, which also mitigate electroweak fine tuning.
}
We introduce $\delta_m$ to show how the constraint relaxes when a tuning is allowed by $m_P^2 = \delta_m \Delta m_P^2$. This model relies on supersymmetry to control quantum corrections to the scalar potential. Accordingly, the constraints on the model parameter space, in particular the upper bound on $m_N$, get stronger as the slepton mass scale $m_{\tilde L}$ is raised.  Direct searches for sleptons at the LHC are thus important for shaping the overall parameter space of the model.%
\footnote{
Note that $m_P$ also receives quantum corrections from the soft mass of the Higgs. Since a large Higgsino mass requires a comparable soft mass for the Higgs in order to obtain the observed electroweak scale, avoiding fine-tuning also requires the Higgsinos not to be too heavy. A detailed consideration of the SM superpartner spectrum in this scenario is beyond the scope of this work, but is an obvious topic for follow-up work.
}
Hereafter, we take $m_{\tilde L} = 100$ GeV, which is consistent with searches for electroweakly-produced superpartners provided neutral lightest supersymmetric particles are similar in mass to charged sleptons  \cite{ATLAS:2024lda, CMS:2024gyw}; collider limits can also be weakened when right- and left-handed sleptons are not mass-degenerate. Larger $m_{\tilde L}$ is viable in this model at the cost of introducing a finer tuning $\delta_m$.

Finally, we comment that in our discussion of the axion's cosmological evolution, we have been neglecting thermal contributions to its mass.  This is a good approximation when $\dot\theta \gtrsim \lambda T$ and 
\begin{equation}
\label{eq:therm_mass}
    \lambda \lesssim 4 \times 10^{-9} ~ c_B^{-1} .
\end{equation}
Although this axion baryogenesis scenario may still be viable in the presence of a large thermal contribution to the axion mass, a detailed analysis of this case is beyond the scope of this work.

The constraints in Eqs.~(\ref{eq:eq_con}) and (\ref{eq:quantum_correct}) give a lower bound on $m_P$:
\begin{equation}
\label{eq:mP_min_tuned}
    m_P \gtrsim 0.8 ~{\rm eV}
    ~ \delta_m \left( \frac{c_{B-L}}{c_B} \right)
    \left( \frac{m_{\tilde L}}{100 ~{\rm GeV}} \right)^2
    \left(\frac{\ln(\Lambda/m_{\tilde L})}{6.9} \right)^2 .
\end{equation}
This is in very mild tension with Eq.~(\ref{eq:mP_max}), which then necessitates a tuning at the level of 
\begin{equation}
\label{eq:tuning_low_mN}
    \delta_m \lesssim 18 \% 
    \left( \frac{c_B}{c_{B-L}} \right)
    \left( \frac{100 ~{\rm GeV}}{m_{\tilde L}} \right)^2
    \left(\frac{6.9}{\ln(\Lambda/m_{\tilde L})} \right)^2 .
\end{equation}
The constraint on $m_P$ in Eq.~(\ref{eq:mP_min_tuned}) together with Eqs.~(\ref{eq:eq_con}) and (\ref{eq:therm_mass}) give a lower bound on $y$:
\begin{equation}
\label{eq:y_min_low_mN}
    y \gtrsim 9 \times 10^{-3} 
    \left( \frac{ c_B c_{B-L} \delta_m}{18\%} \right)^{\scalebox{1.01}{$\frac{1}{2}$} }
    \left( \frac{m_{\tilde L}}{100 ~{\rm GeV}} \right) \left(\frac{\ln(\Lambda/m_{\tilde L})}{6.9} \right).
\end{equation}
Although $y$ can be as large as unity, the allowed values of $\lambda$ and $f_a$ are too small to explain the SM neutrino masses ($m_\nu \sim \lambda y^2 f_a m_N^2 / v^2$), and an additional neutrino mass mechanism needs to be invoked to explain the observed magnitude of the neutrino mass splitting of $\mathcal{O}(0.1)~{\rm eV}$.

\begin{figure}[b]
\includegraphics[width=\columnwidth]{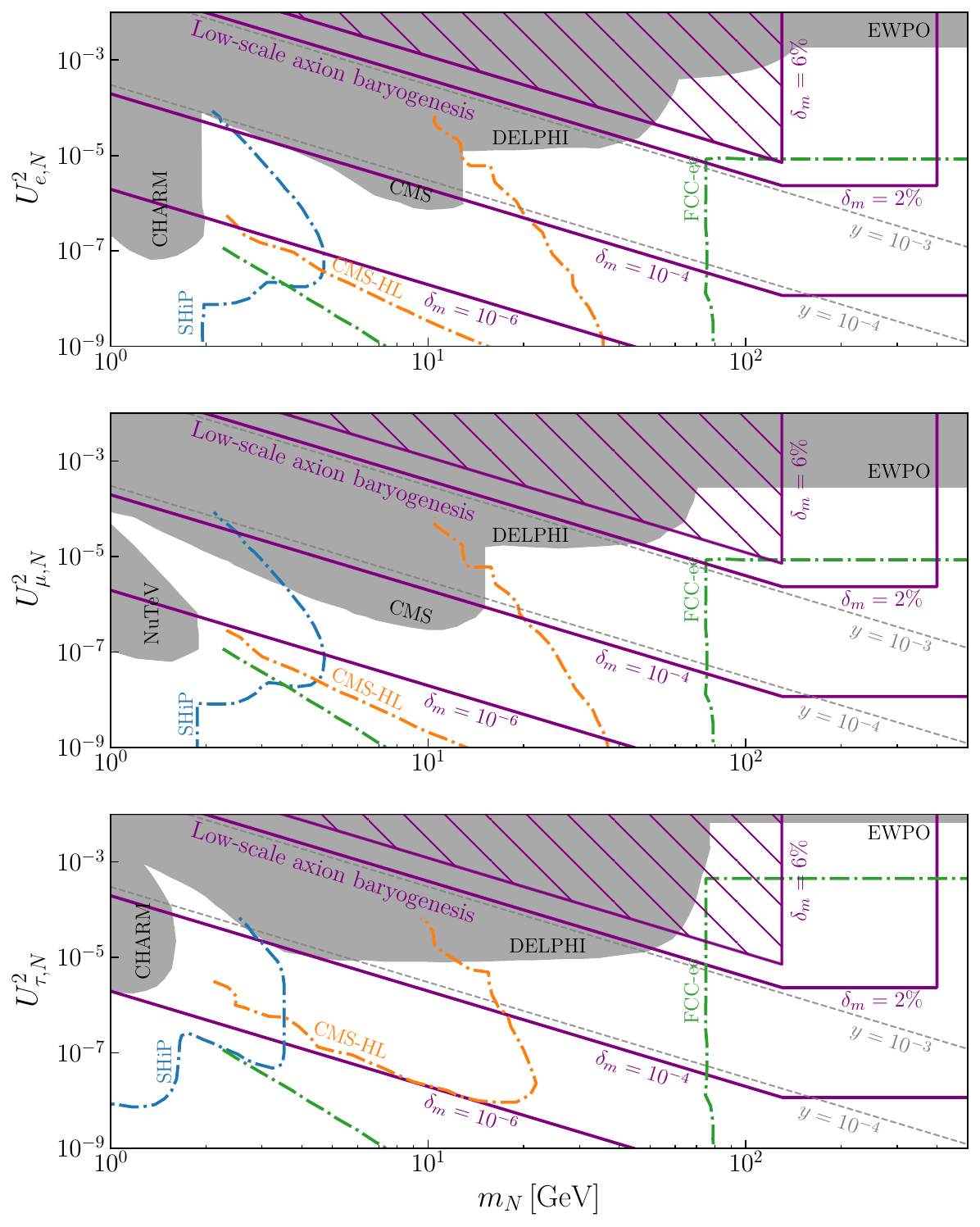}
\caption{Constraints from mixing of $N$ with $e$, $\mu$, and $\tau$ neutrinos. The vertical purple line segments show the upper bound on $m_N$ based on Eqs.~(\ref{eq:tuning_low_mN}) and (\ref{eq:tuning_high_mN}) with slepton mass scale $m_{\tilde L} = 100 \,{\rm GeV}$. The sloped purple segments show the lower bound on the mixing angle derived from Eqs.~(\ref{eq:y_min_low_mN}) and (\ref{eq:y_min_high_mN}). We set $c_B$ and $c_{B-L}$ for one generation of $N$. Different purple lines result from different amounts of fine-tuning $\delta_m$, as labeled, in the radial mode mass. 
Dark gray shaded regions show the existing constraints on the parameter space. Dot-dashed lines show projected 95\% CL sensitivity at the FCC-ee (green), HL-LHC (orange) and SHiP (blue). As can be seen, the most natural region of the viable parameter space is accessible at the FCC-ee, while the parameter space accessible to HL-LHC and SHiP involves a higher degree of fine-tuning.}
\label{fig:mixing}
\end{figure}

The relatively light sterile neutrinos mix with SM neutrinos with a mixing angle of $ \theta_\nu = \frac{y v}{\sqrt{2} m_{N}} = U_{\nu,N}$, where $\nu = (e,\mu,\tau)$ denotes the flavor index of the active neutrino. This mixing gives rise to potentially observable signatures in current and future accelerator-based experiments. We determine these constraints by assuming for simplicity that the sterile neutrinos mix dominantly with a single flavor of SM neutrino\footnote{The assumption of single-flavor alignment is not fine-tuned since our SM neutrino masses do not arise from active neutrinos coupling to sterile neutrinos $N$. Radiative corrections away from the alignment limit are small because of the small active neutrino masses.}, and show the resulting parameter space for mixing with electron, muon, and tau neutrinos in the three panels of Fig.~\ref{fig:mixing}.
The purple sloped line segments at $m_N < T_{\rm EW} = 130 ~{\rm GeV}$ are determined by Eq.~(\ref{eq:y_min_low_mN}) with both $c_B$ and $c_{B-L}$ set for one generation of $N$. The segments at $m_N > 130$~GeV will be discussed at the end of this section.
%
%%%%%%%%%%%%%%%%%%%%%%%%%%%%%%%%%%%%%%%%%
%%%%%%%%%%%%%%%%%%%%%%%%%%%%%%%%%%%%%%%%%

The present constraints on these mixing angles for sterile neutrino masses between $1$-$500 \,{\rm GeV}$ are shown in dark gray and include constraints from CHARM \cite{CHARM:1985anb, Boiarska:2021yho}, BELLE \cite{Belle:2013ytx}, NuTeV \cite{NuTeV:1999kej}, CMS \cite{CMS:2022fut} and DELPHI \cite{DELPHI:1996qcc} experiments. The current bounds on the mixing angles from indirect constraints and electroweak precision observables (EWPO) are calculated with a MCMC following the prescription given in \cite{Antusch:2014woa,Antusch:2015mia} with input quantities updated to the current values tabulated in \cite{Mannel:2020fts, Workman:2022ynf}. We find the following $95\%$  bounds on mixing angles in the alignment limit: $U^{2}_{e,N} < 2\times 10^{-3}$, $U^{2}_{\mu,N} < 3\times 10^{-4}$ and $U^{2}_{\tau,N} < 7\times 10^{-3}$.\footnote{Like Refs.~\cite{Antusch:2014woa,Antusch:2015mia}, we find a $\simeq 2\sigma$ preference for non-zero $U^{2}_{e,N}$ and $U^{2}_{\tau_N}$. However here we only report the $95\%$ upper bound on the mixing angles.}

In Fig.~\ref{fig:mixing}, we also show the projected exclusion sensitivities for searches at a future lepton collider (FCC-ee) \cite{Antusch:2015mia}. This sensitivity results from a combination of displaced vertex and $Z$-pole searches, which dominate the reach for for $m_N < m_Z/2$, as well as improvements to EWPO constraints that control the sensitivity for $m_N> m_Z/2$. We also show the projected sensitivity for displaced  searches at CMS in the HL-LHC run \cite{Drewes:2019fou} and for SHiP (assuming $2\times 10^{20}$ POT) \cite{SHiP:2018xqw}. For a recent summary of constraints on sterile neutrino mixing angles for a given mass, see~\cite{Abdullahi:2022jlv}.

As evident in the figure, the bulk of the natural parameter space in this model of low-scale axion baryogenesis is accessible at the FCC-ee, particularly through the improved precision of EWPO. The lighter and more weakly-coupled right-handed neutrinos that are accessible at HL-LHC and SHiP require a large amount of fine-tuning in the potential for the radial mode.

The bounds in Fig.~\ref{fig:mixing}, including current and predicted EWPO constraints, are computed under the assumption that heavy neutral leptons are the only relevant BSM state; the slepton mass scale enters only through the indicated levels of fine-tuning.  To minimize fine-tuning in the radial mode potential, however, the sleptons should also be relatively light. Thus, the collider signatures, and in particular the sensitivity of EWPOs, may be enhanced depending on the specific implementation of the supersymmetric extension of the SM.

We now comment on the consequences of relaxing some of the assumptions in the previous discussion. If the equilibrium condition imposed in Eq.~(\ref{eq:eq_con}) is not met, the baryon asymmetry is suppressed by the ``freeze-in'' factor $\Gamma_N / H$ with $\Gamma_N$ shown in Eq.~(\ref{eq:rate}), giving 
\begin{equation}
\label{eq:YB_FI}
    Y_B = \left. \frac{c_B \dot\theta T^2}{s} \frac{\Gamma_N}{H} \right|_{T = T_{\rm EW}}.
\end{equation}
Then $Y_B \propto \dot\theta r^2$, which is simply the charge density and therefore independent of the transition temperature $T_q$. This implies that, once the energy density is fixed at recombination for the Hubble tension, there is no additional free parameter that can  open up parameter space beyond Eq.~(\ref{eq:eq_con}). In other words, requiring $Y_B$ in Eq.~(\ref{eq:YB_FI}) to reproduce the observed baryon asymmetry simply saturates the inequality in Eq.~(\ref{eq:eq_con}). 

Thus far, we have assumed $m_N < 130$ GeV so that $N$ remains in the thermal bath at the electroweak phase transition. In the opposite case, when $N$ falls out of thermal equilibrium at $T \simeq m_N > T_\mathrm{EW}$, the $B-L$ asymmetry freezes out and gets preserved through the electroweak phase transition, resulting in a final baryon asymmetry given by~\cite{Harvey:1990qw}
\begin{equation}
\label{eq:YB_FO}
    Y_B = \frac{28}{79} Y_{B-L} = \left. \frac{28}{79} \frac{c_{B-L} \dot\theta T^2}{s} \right|_{T = m_N}.
\end{equation}
In the present model, $c_{B-L}$ is given below Eq.~(\ref{eq:rate}). When the quartic potential dominates as we have assumed, $\dot\theta$ scales linearly with temperature. Therefore, Eq.~(\ref{eq:YB_FO}) is independent of the temperature at which it is evaluated and gives a similar result as Eq.~(\ref{eq:rate}). However, the equilibrium condition in Eq.~(\ref{eq:eq_con}) now has to be imposed at the freeze-out temperature $T = m_N$, which leads to the constraint
\begin{equation}
\label{eq:eq_con2}
    \lambda y \gtrsim 8 \times 10^{-11}
    \left( \frac{m_P}{0.1 ~{\rm eV}} \right)^{\scalebox{1.01}{$\frac{1}{2}$} }
    \left( \frac{m_N}{300 ~{\rm GeV}} \right)^{\scalebox{1.01}{$\frac{1}{2}$} } ,
\end{equation}
which, together with the constraint from mass tuning in Eq.~(\ref{eq:quantum_correct}), results in a lower bound on $m_P$:
\begin{equation}
\label{eq:mP_min_tuned2}
    m_P \gtrsim 5 ~{\rm eV}
    ~ \delta_m
    \left( \frac{m_{\tilde L}}{100 ~{\rm GeV}} \right)^2
    \left(\frac{\ln(\Lambda/m_{\tilde L})}{6.9} \right)^2 
    \left( \frac{m_N}{300 ~{\rm GeV}} \right) .
\end{equation}
In order for $m_P$ to satisfy Eq.~(\ref{eq:mP_max}), there is necessary amount of tuning given by
\begin{equation}
\label{eq:tuning_high_mN}
    \delta_m \lesssim 3\% 
    \left( \frac{300 ~{\rm GeV}}{m_N} \right)
    \left( \frac{100 ~{\rm GeV}}{m_{\tilde L}} \right)^2
    \left(\frac{6.9}{\ln(\Lambda/m_{\tilde L})} \right)^2 .
\end{equation}
This sets the upper bound on $m_N$  (vertical purple segments in Fig.~\ref{fig:mixing}) at $m_N > T_{\rm EW} = 130 ~{\rm GeV}$, where a larger $m_N$ requires a finer tuning.
The constraint on $m_P$ in Eq.~(\ref{eq:mP_min_tuned2}) together with Eqs.~(\ref{eq:eq_con2}) and (\ref{eq:therm_mass}) give a lower bound on $y$:
\begin{align}
\label{eq:y_min_high_mN}
    y \gtrsim 8 \times 10^{-3}
    ~ c_{B-L} &
    \left( \frac{\delta_m}{3\%} \right)^{\scalebox{1.01}{$\frac{1}{2}$} } 
    \left( \frac{m_{\tilde L}}{100 ~{\rm GeV}} \right) \nonumber \\
    \times & \left(\frac{\ln(\Lambda/m_{\tilde L})}{6.9} \right)
    \left( \frac{m_N}{300 ~{\rm GeV}} \right) .
\end{align}
This sets the sloped purple segments in Fig.~\ref{fig:mixing} at $m_N > T_{\rm EW} = 130 ~{\rm GeV}$ with $c_{B-L}$ set for one generation of $N$, where a smaller $y$ requires a finer tuning.

%%%%%%%%%%%%%%%%%%%%%%%%%%%%%%%%%%%%%%%%
\section{Discussion}
\label{sec:discussion}
%%%%%%%%%%%%%%%%%%%%%%%%%%%%%%%%%%%%%%%%

Cosmologies with rotating axions provide an appealing approach to baryogenesis.  They also provide a UV-complete and natural realization of scalar fields with non-trivial time-dependent equations of state.  In this paper we explore the consequences of realizing baryogenesis with a rotating axion that makes a transition from kination to matter during the epoch probed by the CMB.  

On the phenomenological side, axion kination is a qualitatively novel phenomenological mechanism for addressing the Hubble tension: it provides a self-consistent and UV-complete framework that allows for a time-dependent contribution to the effective cold dark matter density during recombination with (i) a sudden disappearance with redshift and (ii) no corresponding time-dependence in an additional dark radiation species. We find that allowing a percent-level fraction of the energy density of the universe to be in the form of a rotating axion during recombination reduces the Hubble tension by more than one sigma.  The presence of axion kination during recombination does not allow for values of $H_0$ as large as can be realized in EDE (see Ref.~\cite{Poulin:2023lkg} and references therein) or stepped self-interacting DM-DR models \cite{Aloni:2021eaq, Schoneberg:2022grr, Joseph:2022jsf, Buen-Abad:2022kgf, Allali:2023zbi, Buen-Abad:2023uva, Schoneberg:2023rnx}, and as such does not provide as significant a resolution to the Hubble tension.  However, axion kination presents a significantly better fit to the data than does $\Lambda$CDM alone, and most importantly does so {\it without} introducing sizeable shifts in either $n_\mathrm{s}$ or $\omega_\mathrm{c}$, thus avoiding exacerbating or introducing tensions in other cosmological datasets.

On the model side, placing the kination-to-matter transition during the CMB epoch means that the overall abundance of the axion is tightly constrained, and cannot exceed a few percent of the total energy density at recombination.  Requiring that this relatively small axion energy density, and thus relatively small PQ charge density, can successfully generate the observed baryon asymmetry of the universe then places several new demands on the cosmic history of the axion.  The initial axion field velocity must be large, which we accomplish by adding additional quartic terms to the axion potential that affect its evolution at early times.
To realize circular rotation, the radial mode of the complex scalar must thermalize through scatterings with a radiation bath \cite{Co:2019wyp}. For the transition from matter to kination to occur post-recombination, this thermalization most easily proceeds through dark fermions $\psi$ that do not interact with the SM. The energy density of these fermions is generically negligible in comparison with that of the rotating axion, but could constitute an interesting extension of the signatures of this model in certain regions of parameter space.  

We transfer the axion PQ charge to SM lepton number via right-handed neutrinos.  Requiring a low-scale rotating axion to successfully transfer a sufficiently large baryon asymmetry to match observations imposes several conditions on the mass and couplings of these sterile neutrinos, singling out a distinctive region of parameter space.  The resulting relatively heavy and strongly-coupled heavy neutral leptons are motivated and achievable targets for future lepton colliders, especially through their imprint on electroweak precision observables.

\vspace{0.5cm}

{\bf Acknowledgments.}---%
We thank Gil Holder and Tristan Smith for useful conversations.
The work of R.C.~was supported in part by DOE grant DE-SC0011842 at the University of Minnesota. 
The work of N.F.~is supported by DOE grant DE-SC0010008. N.F.~thanks the Sloan Foundation for its partial support. The work of A.G. is supported by DOE grant DE–SC0007914. 
The work of K.H.~was supported by Grant-in-Aid for Scientific Research from the Ministry of Education, Culture, Sports, Science, and Technology (MEXT), Japan (20H01895), and by World Premier International Research Center Initiative (WPI), MEXT, Japan (Kavli IPMU).
The work of J.S.~was supported in part by DOE grants DE-SC0023365 and DE-SC0015655. J.S.~gratefully acknowledges MIT's generous hospitality during the performance of this work. 
N.F. and A.G.~thank the Aspen Center for Physics, which is supported by National Science Foundation grant PHY-2210452.

\onecolumngrid
\vspace{1cm}
\appendix
\begin{center}
\textbf{\large Appendix}\\ 

\end{center}

%%%%%%%%%%%%%%%%%%%%%%%%%%%
\section{Axion rotations in the two-field model}
\label{sec:two-field}
%%%%%%%%%%%%%%%%%%%%%%%%%%%%

In this paper, we consider a supersymmetric two-field axion model that realizes a rapid transition from a matter-like to kination-like equation of state, first introduced in~\cite{Co:2021lkc}. In this section we provide a quick review of the physics of this two-field model and then briefly discuss the phenomenological consequences of requiring this model to mitigate the Hubble tension. 

The superpotential of the model is
\begin{align}
W = \lambda X (P \bar{P} - v_{\rm PQ}^2),
\end{align}
where $X$ is a $U(1)$-neutral chiral field, $P$ and $\bar{P}$ are chiral fields with opposite $U(1)$ charges, and $\lambda$ and $v_{\rm PQ}$ are constants. The resulting supersymmetric potential is
\begin{align}
    V = \lambda^2 |P\bar{P}-v_{\rm PQ}^2|^2 + \lambda^2|X|^2 \left( |P|^2 + |\bar{P}|^2 \right).
\end{align}
The first term fixes $P$ and $\bar{P}$ on the moduli space $P\bar{P}=v_{\rm PQ}^2$, and the second term fixes $X$ at $X=0$. We may then integrate out $X$ and a linear combination of $P$ and $\bar{P}$ via these two relations, leaving one complex scalar degree of freedom. Without loss of generality, we choose $P$ as a low-energy degree of freedom. Its kinetic term is non-canonical,
\begin{align}
- \partial_\mu P^\dag \partial^\mu P  - \partial_\mu \bar{P}^\dag \partial^\mu \bar{P} \rightarrow - \left( 1 + \frac{v_{\rm PQ}^4}{|P|^4}\right) \partial_\mu P^\dag \partial^\mu P.
\end{align}
While $P$ does not have a potential in the supersymmetric limit, soft supersymmetry breaking mass terms $ m_P^2 |P|^2 + m_{\bar{P}}^2 |\bar{P}|^2$ for $P$ and $\bar{P}$ give a non-zero potential,
\begin{align}
\label{eq:two-field}
  V(P) = \left( 1 + \frac{m_{\bar{P}}^2}{m_P^2} \frac{v_{\rm PQ}^4}{|P|^4} \right) m_P^2|P|^2 \equiv \left( 1 + r_P^2 \frac{v_{\rm PQ}^4}{|P|^4} \right) m_P^2|P|^2 ,
\end{align}
where we introduce the parameter $r_P \equiv m_{\bar{P}} / m_P$.  The equation of state of the axion rotation derived in the next section will depend on $r_P$, and thus it  will ultimately govern the exact transition from matter-like to kination-like.
The effective Lagrangian is given by
\begin{align}
\label{eq:Leff}
    {\cal L}_{\rm eff} = - \left( 1 + \frac{v_{\rm PQ}^4}{|P|^4}\right) \partial_\mu P^\dag \partial^\mu P -  \left( 1 + r_P^2 \frac{v_{\rm PQ}^4}{|P|^4} \right) m_P^2|P|^2 .
\end{align}
The value of $|P|$ at the minimum of the potential is given by $|P| = \sqrt{r_P} v_{PQ}$. We assume $r_P \gtrsim 1$.
For $|P| \gg v_{\rm PQ}$, the effective Lagrangian approaches that of a free massive scalar with a canonical kinetic term.

Axion rotations may be initiated by the Affleck-Dine mechanism~\cite{Affleck:1984fy}. In the early universe, the coupling of $P$ with gravity generically introduces a potential for $P$ that is proportional to the total energy density of the universe. The dominant term is the term quadratic in $P$, which is called the Hubble-induced mass:
\begin{align}
    V_H = c H^2 |P|^2.
\end{align}
If $c < 0$, $P$ is driven to a large field value $\gg v_{\rm PQ}$. For large field values of $P$, higher-order terms in $P$ may be important, and some of them may explicitly break the $U(1)$ symmetry.

We first discuss the simplest case where the explicit breaking is given by a single term in the superpotential of the form
\begin{align}
\label{eq:W U(1)breaking}
    W = \frac{P^{n+1}}{(n+1)M^{n-2}},
\end{align}
where $M$ is a cutoff scale.
This superpotential gives rise to both supersymmetric and supersymmetry-breaking contributions to the potential,
\begin{align}
\label{eq:U1breaking}
V = \frac{1}{M^{2n-4}}|P|^{2n} + \left(\frac{A}{M^{n-2}} P^{n+1} + {\rm h.c.} \right),
\end{align}
where $A$ is a soft supersymmetry- and $R$- breaking parameter whose natural magnitude is $m_P$. The first term preserves the $U(1)$ symmetry, and stabilizes the radial direction of $P$ against the negative Hubble-induced mass term, so that $|P|$ follows an attractor solution
\begin{align}
    |P| \sim H^{\frac{1}{n-1}}M^{\frac{n-2}{n-1}}.
\end{align}
for $H \gg m_P$~\cite{Dine:1995kz,Harigaya:2015hha}.  Around $H \sim m_P$, $P$ no longer follows the attractor solution and begins to oscillate around the origin driven by the supersymmetry-breaking mass term $m_P$. At the same time, the second term in Eq.~(\ref{eq:U1breaking}), which explicitly breaks the $U(1)$ symmetry, kicks $P$ in the angular direction, and $P$ rotates around the origin.
One can show that the potential gradients to the radial and angular directions are of the same order when $H\sim m_P$. This guarantees that the energy densities of the radial and angular modes are of the same order.
Note that the explicit $U(1)$ breaking term is proportional to supersymmetry and R symmetry breaking, since the superpotential alone preserves a linear combination of the $U(1)$ symmetry of $P$ and the $R$ symmetry.
The field value of $|P|$ decreases in proportion to $a^{-3/2}$, and the higher-order $U(1)$ breaking term becomes inefficient, so that $P$ continues to rotate while approximately preserving the angular momentum in field space up to dilution by the cosmic expansion.

We note that, for the purpose of addressing the Hubble tension discussed in Secs.~\ref{sec:axion_cosmo} and~\ref{sec:fit_to_data}, we only need to utilize the nearly quadratic potential  and the simplest explicit $U(1)$ breaking described above. However, when discussing the connection to the baryon asymmetry in Sec.~\ref{sec:baryons}, we assume an additional quartic term that dominates at high temperatures in order to generate sufficient baryon asymmetry at the electroweak phase transition. In supersymmetric theories, such a quartic term can be simply generated by a superpotential term $W \supset \lambda_r^{1/2} Z P^2$ with $Z$ a chiral field.

When this extra quartic potential is present, the Affleck-Dine mechanism utilizing the explicit $U(1)$ breaking in Eq.~\eqref{eq:W U(1)breaking} is not efficient. This is because the potential gradient in the radial direction is much larger than that in the angular direction. Consequently, the energy density of the radial mode $\rho_r$ dominates over that of the angular mode $\rho_{\rm PQ}$ by a factor of $\epsilon$, which is given by the ratio of the potential gradients in the angular to the radial modes. This large radial energy density can lead to an over-production of dark radiation when $\rho_r$ is thermalized into a dark sector thermal bath. 

We now analyze the size of $\epsilon$ necessary to generate sufficient $Y_{\rm PQ}$ in the context of Hubble tension and baryogenesis, while not over-producing dark radiation. To address the Hubble tension, $\left. \rho_{\rm PQ} \right|_{\rm rec} = m_P^2 v_{\rm PQ}^2$ should be around $1\%$ of the total energy density at $a=a_c$, $ \rho_{\rm tot} (a_c) \sim 0.026~{\rm eV}^4$, which also fixes the charge yield $Y_{\rm PQ} = n_{\rm PQ} / s(T_{\rm rec})$. The quartic coupling $\lambda_r$ is then fixed from matching the quadratic and quartic potential energy at the temperature $T_q$ given in Eq.~(\ref{eq:T_q}), $\rho_{\rm PQ} (T_q / T_{\rm rec})^3 = \lambda_r r(T_q)^4 /4$ with $r(T_q) = v_{\rm PQ} \left[g_{*S}(T_q) T_q^3 / (g_{*S}(T_{\rm rec}) T_{\rm rec}^3) \right]^{1/2}$. Assuming an initial radial field value of $r_i$ when the complex field starts the rotation, the yield of the radial mode number density $Y_r = n_r/s$ is computed by $\sqrt{\lambda_r} r_i^3 / s(T_{\rm osc})$ with $T_{\rm osc}$ obtained from the oscillation condition $3H(T_{\rm osc}) \sim \sqrt{\lambda_r} r_i$. The temperature $T_M$ at which $\rho_r$ dominates is given by $m_P Y_r s(T_M) = \pi^2 g_* T_M^4/30$, while the radial field value at this time is given by $r_M \equiv r(T_M) = v_{\rm PQ} (T_M / T_{\rm rec})^{3/2}$.  
We define $\epsilon$ as the radio of the angular and radial yields $\epsilon = Y_{\rm PQ}/Y_r$. With all the relations listed above, we find 
\begin{align}
    r_i & \simeq 2 \times 10^{17} ~{\rm GeV} 
    \left( \frac{0.1~{\rm eV}}{m_P} \right)^{\scalebox{1.01}{$\frac{1}{2}$} }
    \left( \frac{10^{-6}}{\epsilon} \right)^{\scalebox{1.01}{$\frac{2}{3}$} }
    \left( \frac{21/257}{c_B} \right)^{\scalebox{1.01}{$\frac{1}{2}$} }
    , \nonumber \\
    r_M & \simeq 10 ~{\rm MeV} 
    \left( \frac{0.1~{\rm eV}}{m_P} \right)
    \left( \frac{10^{-6}}{\epsilon} \right)^2 , \\
    T_M & \simeq 400 ~{\rm eV}
    \left( \frac{10^{-6}}{\epsilon} \right) .
\end{align}
This demonstrates that there is viable parameter space where $r_i < M_{\rm Pl}$ for $m_P \lesssim \left( \left. \rho_{\rm PQ} \right|_{\rm rec} \right)^{1/4} \sim 0.1\, {\rm eV}$ with $\epsilon < 1$. 
Finally, the thermalization of the complex field can proceed via a Yukawa coupling with a dark fermion, $\mathcal{L} \supset y_P P \psi \bar\psi$. In order for $\psi, \bar\psi$ to be in thermal equilibrium at temperature $T$, $y_P r(T) \lesssim T$ and consequently thermalization processes have a maximum rate $\Gamma_{\rm th} \simeq 0.1 y_P^2 T \lesssim 0.1 T^3/r^2(T)$. Ensuring that thermalization proceeds before $\rho_r$ dominates, we require $\Gamma_{\rm th}(T_M) \ge H(T_M)$, and thus $0.1 T_M^3/r^2(T_M) \ge H(T_M)$. This constraint turns out to be less stringent than the combination of the aforementioned constraints $r_i < M_{\rm Pl}$ for $m_P \lesssim {\rm eV}$.

However, if we had set $\epsilon = m_P / (\sqrt{\lambda_r} r_i)$ as would be the case from Eq.~(\ref{eq:U1breaking}), we would find no consistency in the parameters because the expected $\epsilon$ is too small to keep $m_P \lesssim {\rm eV}$.
Note that suppression of $\epsilon$, which is proportional to $A\sim m_P$, stems from the $R$ symmetry in the superpotential \eqref{eq:W U(1)breaking}. 
Therefore, the issue of too small $\epsilon$ can be remedied by allowing for additional $R$-symmetry breaking, which then generates a stronger kick than the $A$-term in the potential. Specifically, we may assume a superpotential
\begin{equation}
    W = \sqrt{\lambda_r} Z P^2 + \frac{P^{\widetilde{n}+1}}{(\widetilde{n}+1)\tilde{M}^{\widetilde{n}-2}} + \frac{P^{n+1}}{(n+1)M^{n-2}},
\end{equation}
where the different powers of $P$ result in $R$-symmetry breaking.
The potential of $P$ is
\begin{equation}
\label{eq:Lag_R_symm}
    \mathcal{L} \supset \lambda_r |P|^4 + V(M,n) + V(\tilde M, \widetilde{n}) + \left(\frac{P^{\widetilde{n}} P^n}{\tilde{M}^{\widetilde{n}-2} M^{n-2}} + {\rm h.c.}\right)
\end{equation}
plus that in Eq.~\eqref{eq:Leff},
where $V(M, n)$ is defined as the potential terms in Eq.~(\ref{eq:U1breaking}) and $V(\tilde M, \widetilde{n})$ is the analogous version.
\if0
The first term in Eq.~(\ref{eq:Lag_R_symm}) provides the quartic potential, while the $A$-term in $V(\tilde M, \widetilde{n})$ provides a stronger kick than that in Eq.~(\ref{eq:U1breaking}) because ${\widetilde n}$ can be smaller than $n$ \RC{is this explanation correct?}.
\fi
The first term in Eq.~(\ref{eq:Lag_R_symm}) provides the quartic potential, while the fourth term provides a stronger kick than that in Eq.~(\ref{eq:U1breaking}).
This extra assumption in the model simply allows $\epsilon$ to become a free parameter and does not interfere with the earlier discussions of the Hubble tension and baryogenesis.

%%%%%%%%%%%%%%%%%%%%%%%%%%%
\section{Boltzmann equations for a rotating axion}
\label{sec:Pert_eqs}
%%%%%%%%%%%%%%%%%%%%%%%%%%%%

In this section we provide a detailed derivation of the equations governing the evolution of the homogeneous rotating axion background as well as the perturbations for a general axion potential $V(r)$.
We are interested in cosmologies where a complex scalar field $P = 1/\sqrt{2} \, r e^{i\theta}$ undergoes rapid rotations,  $\dot{\theta} \gg H$, and study the perturbations of this system at times when the angular mode $\theta$ is the only light degree of freedom.  In this regime, the potential $V=V(r)$ is invariant under the $U(1)$ symmetry taking $P\to e^{i\alpha}P$, and the kinetic terms for the radial mode $r$ in the Lagrangian can be neglected.
This means we will be able to algebraically solve for both the background $r_0$ and the perturbation $\delta r$ in terms of the dynamical degree of freedom $\theta$.

Since scalar fields do not support anisotropic stress (and we neglect the small contribution from SM neutrinos), we can derive the Boltzmann equations for kinaton perturbations in either synchronous or conformal Newtonian gauge using a metric of the general form
\begin{align}
ds^2 = a^2\left( - \left(1+2A\right)d\eta^2 + \left(1 + 2 D\right) \delta_{ij}dx^idx^j  \right) .
\end{align}
We write $\theta$ in terms of background and perturbation as $\theta \equiv \theta_0 + \theta_1$. For a scalar with a canonical kinetic term undergoing circular rotation in a potential $V(r)$,  the equations governing the background field evolution are
\begin{eqnarray}
\label{eq:bkgd}
\partial_{r_0} V (r_0) &=& r_0 \frac{1}{a^2} \theta_0'^2\\
\label{eq:bkgdtheta}
(\partial_\eta + 3\cH )\left(\frac{1}{a} r_0^2 \theta_0' \right)& =& 0,
\end{eqnarray}
where a prime indicates a derivative with respect to conformal time $\eta$ and $\cH \equiv a'/a$. The homogeneous energy density $\bar\rho$ and pressure $\bar P$ are then
\beq
\bar \rho =  \frac{1}{2a^2} r_0^2 \theta_0'^2 + V (r_0), \phantom{space} \bar P =  \frac{1}{2a^2} r_0^2 \theta_0'^2 - V (r_0).
\eeq
The equation of state parameter is, using Eq.~(\ref{eq:bkgd}),
\beq
\label{eq:wdef}
w = \frac{r_0 \partial_{r_0} V(r_0)-2 V}{r_0 \partial_{r_0}V(r_0)+2 V}.
\eeq
Eqs.~(\ref{eq:bkgd}) and~(\ref{eq:bkgdtheta}) also let us relate the time evolution of $r_0$ to that of $\theta_0'$, giving
\begin{eqnarray}
\label{eq:bkgd1}
\frac{r_0'}{r_0}  &=& -3 \cH \frac{f}{1+2f} \\
\label{eq:bkgd2}
\frac{\theta_0''}{\theta_0'} &=& -2 \cH \frac{1-f}{1+2f},
\end{eqnarray}
where the function $f$ encapsulates the dependence on the potential:
\beq
\label{eq:fdef}
f  \equiv  \frac{2 \partial_{r_0}V(r_0)}{r_0 \partial_{r_0}^2 V(r_0) -  \partial_{r_0} V(r_0)}.
\eeq
Taking the time derivative of the equation of state, 
\beq
w' = 3\cH (1-w)\left(w- \frac{1}{1+2f}\right),
\eeq
allows us to exchange $f$ for $w'$. 

Writing $r=r_0+\delta r$, $\theta = \theta_0 + \delta\theta$, the equation of motion for $r$ gives at leading order in perturbations
\beq
\label{eq:deltar}
\frac{\delta r}{r_0}  = f  \left(  \frac{\theta_1}{\theta_0'} -A\right),
\eeq
where again all dependence on the potential enters through $f$. 
Using Eqs.~(\ref{eq:bkgd1}),~(\ref{eq:bkgd2}), and~(\ref{eq:deltar}), we can express the perturbations $\delta\rho$ and $\delta P$ in terms of  $\theta$, the metric perturbation $A$, and $f$:
\begin{eqnarray}
    \label{eq:rhopert}
\delta\rho&=& \bar \rho \left( \frac{\theta_1'}{\theta_0'}-A\right) (1+2f)(1+w),\\
  \label{eq:ppert}
\delta P&=& \bar \rho \left( \frac{\theta_1'}{\theta_0'}-A\right) (1+w).
\end{eqnarray}
We can then read off the speed of sound in this frame,
\beq
\label{eq:cs2}
 c_s^2= \frac{\delta P}{\delta \rho} = \frac{1}{1+2f} = w - \frac{w'}{3\cH (1+w)},
\eeq
which is identical to the adiabatic speed of sound. Note that this is not the case for a canonical single scalar field.
The Boltzmann equation for $\delta \equiv \delta\rho/\rho$ is given in terms of $w(\eta)$ as
\beq
\label{eq:delta}
\delta' + (1+w) (\Theta + 3 D')  -\frac{w'}{1 +w}\delta = 0 . 
\eeq
Similarly, the Boltzmann equation for $\Theta \equiv \partial_i v^i = -\partial_i^2\delta\theta/\theta_0' $ is given in terms of $w(\eta)$ as
\beq
\label{eq:velocity}
\Theta' + \cH(1-3w) \Theta + \frac{w'}{1 +w} \Theta-\frac{c_s^2}{1+w} k^2\delta + k^2 A = 0 .
\eeq
Here as usual $k$ indicates the comoving wave number of the perturbation.
Eqs.~\eqref{eq:delta} and \eqref{eq:velocity} are simply the Boltzmann equations of a generic fluid with time-dependent $w$.

\section{Equation of state in the two-field model}
\label{sec:eos}

In this work we consider the ``two-field'' model of axion kination introduced in \cite{Co:2021lkc}, where the cosmological epoch of interest is governed by the Lagrangian in Eq.~\eqref{eq:Leff}. Here we determine the equation of state $w(a)$ for this model.

We define a new radial variable 
\begin{align}
\label{eq:Fdef}
    F^2 = 2 r^2 \left( 1 + \frac{v_{\rm PQ}^4}{r^4} \right) \equiv 2 r^2 \mathcal{K}
\end{align}
so that the kinetic term is canonical up to terms proportional to $\partial_\mu r$, which can again be neglected in comparison to the dominant contributions proportional to $r \partial_\mu \theta$. In terms of the effective radial degree of freedom $F$, the Lagrangian now reads
\begin{align}
    {\cal L} & \simeq 
    \frac{1}{2} F^2 \left(\partial_\mu \theta \partial^\mu\theta\right) - \frac{1}{4} m_P^2 F^2 \left(  \left(1+r_P^2\right) + \left( 1-r_P^2 \right) \sqrt{1 - \left( \frac{2 v_{\rm PQ}}{F} \right)^4 }   \right). \nonumber
\end{align}
Then the equation of state is given by Eq.~(\ref{eq:wdef}),
\begin{align}
\label{eq:w_F}
    w & = \frac{ 
    \left(\frac{ 2 v_{\rm PQ}}{F} \right)^4 (1 - r_P^2) + 2 r_P \left(\frac{ 2 v_{\rm PQ}}{F} \right)^2 \sqrt{1 - \left(\frac{ 2 v_{\rm PQ}}{F} \right)^4 }
    }{
    2(1-r_P^2) - \left(\frac{ 2 v_{\rm PQ}}{F} \right)^4 (1 - r_P^2) + 2 \left(1 + r_P^2 - r_P \left(\frac{ 2 v_{\rm PQ}}{F} \right)^2 \right) \sqrt{1 - \left(\frac{ 2 v_{\rm PQ}}{F} \right)^4 } } ,
\end{align}
while $f$, defined in Eq.~(\ref{eq:fdef}), is given by
\begin{align}
    f = \frac{ -\left(1-\left(\frac{ 2 v_{\rm PQ}}{F} \right)^4\right) \left( 1 - r_P^2 + (1+r_P^2) \sqrt{1 - \left(\frac{ 2 v_{\rm PQ}}{F} \right)^4 } \right) }{ \left(\frac{ 2 v_{\rm PQ}}{F} \right)^4 (1-r_P^2)} .
\end{align}

Still, one needs to derive the evolution of $F$ with scale factor in order to obtain the evolution of the equation of state. The evolution of $F$ can be explicitly derived using the equations of motion, where Eq.~(\ref{eq:bkgd}) determines the rotational speed $\dot\theta$ for a given effective radial coordinate $F$ and, based on Eq.~(\ref{eq:bkgdtheta}), the conserved $U(1)$ charge yield, defined as the charge density divided by the entropy density, can be expressed in terms of $F$ as $Y_{\rm PQ} \equiv \dot\theta F^2 / \left( 2\pi^2 g_{*S}(T) T^3/45 \right)$. These two equations give $F$ as a function of $T$ for a fixed charge yield $Y_{\rm PQ}$.
It is convenient to express $F$ at an arbitrary temperature in terms of its value at the minimum of the potential, $F_{\rm min} = \sqrt{2(r_P+r_P^{-1})} v_{\rm PQ}$.  For $r_P \to 1$, this minimum is attained, i.e., $F = F_{\rm min}$, at a specific temperature $T_S$. Equivalently, $T_S$ is the temperature at which $w(T) = 1$ in the limit when $r_P$ approaches to unity from above. For an arbitrary $r_P > 1$, $w(T)$ only approaches unity asymptotically, and likewise for $F \rightarrow F_{\rm min}$. For convenience, we define the temperature $T_S$ in this case by assuming that $F \simeq \mathcal{O}(1) F_{\rm min}$ when $T = T_S$. Specifying the precise $\mathcal{O}(1)$ factor $F/F_{\rm min}$ used in the definition becomes somewhat arbitrary for $r_P > 1$; we define it so that $T_S$ reproduces the expression for $r_P \rightarrow 1$ when $r_P$ is set to unity. Following this prescription, we define the temperature $T_S$ for any $r_P$ as
\begin{align}
T_S \equiv \left( \frac{90}{\pi^2g_{*S}} \frac{r_P m_P v_{\rm PQ}^2}{Y_{\rm PQ}} \right)^{\scalebox{1.01}{$\frac{1}{3}$} } ,
\end{align}
where the dependence on $r_P$ is analytically derived by using the prescriptions above and taking the $r_P \gg 1$ limit.

For implementation in CLASS, it is convenient to express evolution of $F$ in terms of the scale factor $a$. In terms of $a_S$, the scale factor at which $T = T_S$, entropy conservation implies $T/T_S = a_S/a$ and the evolution of $F(a)$ can accordingly be written as
\begin{align}
\label{eq:F_Fmin}
\frac{F(a)}{F_{\rm min}} = & \frac{1}{(3 \chi )^{\frac{1}{4}} \left(1+r_P^2\right)^{\frac{1}{2}} }\left[\chi^{\frac{4}{3}}+\chi\left(1+r_P^2\right)\left(1+4 r_P^2 \left(\frac{a_S}{a}\right)^6+r_P^2\right) \right. \\ \nonumber
& + \chi^{\frac{2}{3}}\left(\left(1+r_P^2\right)^4+ \left. 8 r_P^2  \left(\frac{a_S}{a}\right)^6\left(1+r_P^2\right)\left(1-4 r_P^2+r_P^4\right)+16 r_P^4 \left(\frac{a_S}{a}\right)^{12}\left(1-r_P^2+r_P^4\right)\right)\right]^{\frac{1}{4}} , 
\end{align}
where
\begin{align}
\label{eq:F_chi}
\chi \equiv & \left(1+r_P^2\right)^6+32 r_P^6 \left(\frac{a_S}{a}\right)^{18}\left(-2+r_P^2\right)\left(1+r_P^2\right)\left(-1+2 r_P^2\right) \\ \nonumber
& + 12 r_P^2 \left(\frac{a_S}{a}\right)^6\left(1+r_P^2\right)^3\left(1-4 r_P^2+r_P^4\right) + 24 r_P^4 \left(\frac{a_S}{a}\right)^{12}\left(2-7 r_P^2+18 r_P^4-7 r_P^6+2 r_P^8\right) \\ \nonumber
& +24 r_P^5 \left(\frac{a_S}{a}\right)^9 \left(-1+r_P^2\right)\left[-48 r_P^6 \left(\frac{a_S}{a}\right)^{18}+72 r_P^4 \left(\frac{a_S}{a}\right)^{12}\left(1+r_P^2\right)+6\left(1+r_P^2\right)^3 \right. \\ \nonumber
& \hspace{2 in} \left. + 9 r_P^2 \left(\frac{a_S}{a}\right)^6 \left(-5+r_P^2\right)\left(-1+5 r_P^2\right)\right]^{\frac{1}{2}} .
\end{align}
Now one can compute the equation of state $w(a)$ from Eq.~(\ref{eq:w_F}) using Eqs.~(\ref{eq:F_Fmin}) and (\ref{eq:F_chi}). The results are shown by the solid curves in Fig.~\ref{fig:w_cs} in the paper, where various colors indicate different values of $r_P$. For convenience of comparison, we have shown curves with different $r_P$ but equal values of the reference scale factor $a_{1/3}$, where $w = 1/3$ when $a = a_{1/3}$. The function $w(a)$ asymptotes to the orange curve for $r_P \gg 1$.

The transition from matter to kination can thus be characterized by several closely-related but slightly distinct reference scale factors: $a_S$, $a_{1/3}$, and the scale factor $a_c$ used as a free parameter in our fits to data.  In a radiation-dominated universe, $a_{1/3}$ coincides with $a_c$. However, around matter-radiation equality, $a_{1/3}$ and $a_c$ are slightly mismatched, depending on how $a_c$ compares to $a_{\rm eq}$. Therefore, we treat $a_S$ as a free input parameter and compute $a_c$ numerically.  

%%%%%%%%%%%%%%%%%%%%%%%%%%%
\section{Transfer rate of PQ charge to $L$}
\label{sec:transfer_rate}
%%%%%%%%%%%%%%%%%%%%%%%%%%%

In this appendix, we discuss the derivation of the transfer rate of the PQ charge to the SM lepton asymmetry given in Eq.~(\ref{eq:rate}). We focus on the model with the Lagrangian given in Eq.~(\ref{eq:L_transfer}). The lepton number-violating process is mediated by $P N \leftrightarrow L^\dag H^*$ scattering, where the radial degree of freedom in $P$ is assumed to take a background value of $r$ since $P$ rotates with a fixed radius. The corresponding $L$-violating rate is then given by
\begin{equation}
    \Gamma_L = \frac{1}{64 \pi} \frac{y^2 \lambda^2 r^2}{m_N},
\end{equation}
where we assume that $m_N$ is larger than the thermal mass of $H$ and $L$ so that the $L$ violation is dominated by the decay and inverse decay of $N$.

We now compute the production rate of the $N$ asymmetry before computing that for $B-L$.
This can be obtained by computing the production rate of the $N$ asymmetry as a function of the chemical potentials in the limit $\dot\theta = 0$, and then using detailed balance to restore the dependence on $\dot{\theta}$.
When $\dot{\theta}=0$,
the relevant interactions are $N \leftrightarrow LH$ and $\bar N \leftrightarrow L^\dag H^*$ with the amplitude-squared $|{\cal M}|^2 = y^2 \lambda^2 r^2 / 8$. The production rate of the $N$-number asymmetry is given by
\begin{align}
    \dot{n}_N = \int d\Pi_N d\Pi_L d\Pi_H (2\pi)^4 |\mathcal{M}|^2 \delta^4(p_N-p_L-p_H) \times \left[ f_N (1-f_L) (1+f_H) - f_L f_H (1-f_N) \right] + {\rm h.c}. 
\end{align}
where the phase space distribution factors can be approximated in terms of the chemical potentials as
\begin{align}
    \left[ f_N (1-f_L) (1+f_H) - f_L f_H (1-f_N) \right] \simeq \frac{\mu_N - \mu_L - \mu_H}{T} e^{-E/T}.
\end{align}
For non-zero $\dot{\theta}$, since the chemical potential of the rotating $P$ is $\dot{\theta}$,
we may replace $\mu_N - \mu_L - \mu_H $ with $\mu_N - \mu_L - \mu_H + \dot\theta$ using detailed balance. 
In computing the production rate of the $N$ asymmetry, we are interested in the situation where the initial asymmetries of $N$, $L$, and $H$ are zero. The production rate is then given by
\begin{equation}
    \dot{n}_N = \frac{m_N \dot\theta}{T^2} \Gamma_L n_{N, {\rm eq}} = m_N \dot\theta \, \Gamma_L \frac{3 \, \zeta(3) T}{2 \pi^2} ,
\end{equation}
where the factor of three accounts for three generations of $N$.
Finally, the production rate of $B-L$ is twice that of $N$ given that each $P N \leftrightarrow L^\dag H^*$ interaction changes $L$ by two units. Here we include $N$ number in the lepton number $L$ because the $N$ asymmetry is eventually converted into the SM when $N$ decays to $LH$. To estimate the transfer rate $\Gamma_N$, which is defined through $\dot n_{B-L} \equiv \Gamma_N \times n_{B-L, {\rm eq}} $, we need to derive the equilibrium value of the $B-L$ asymmetry $n_{B-L, {\rm eq}} $ in this model. Analogous to the derivation of $c_B$ outlined below Eq.~(\ref{eq:YB}), we find the equilibrium value of $n_{B-L, {\rm eq}} = c_{B-L} \dot\theta T^2 $, with $c_{B-L} = 1225/3084, 1265/1662, 145/132$ for $1,2,$ and $3$ generations, respectively. The final transfer rate is given in Eq.~(\ref{eq:rate}).

\bibliography{kinationH0}

\end{document}